\documentclass[twocolumn]{aastex631}

\usepackage{silence}
\DeclareRobustCommand{\ion}[2]{\textup{#1\,\textsc{\lowercase{#2}}}}
\usepackage{graphicx}
\usepackage{savesym}
\savesymbol{tablenum}
\usepackage{siunitx}
\restoresymbol{SIX}{tablenum}
\usepackage{natbib}

\usepackage{amsmath}
\usepackage{txfonts}
\usepackage{multirow}
\usepackage{float}
\begin{document}

\title{The distance of quasar outflows from the central source: The first consistent values from emission and absorption determinations.}

\author[0009-0001-5990-5790]{Mayank Sharma}
\affiliation{Department of Physics, Virginia Tech, Blacksburg, VA 24061, USA}
\email{mayanksh@vt.edu}

\author[0000-0003-2991-4618]{Nahum Arav}
\affiliation{Department of Physics, Virginia Tech, Blacksburg, VA 24061, USA}

\author{Qinyuan Zhao}
\affiliation{CAS Key Laboratory for Research in Galaxies and Cosmology, Department of Astronomy, University of Science and Technology of China, Hefei 230026, China}
\affiliation{School of Astronomy and Space Sciences, University of Science and Technology of China, Hefei 230026, China}
\affiliation{Department of Astronomy, Xiamen University, Xiamen, Fujian 361005, China}

\author[0000-0002-0964-7500]{Maryam Dehghanian}
\affiliation{Department of Physics, Virginia Tech, Blacksburg, VA 24061, USA}

\author[0000-0002-3687-6552]{Doyee Byun}
\affiliation{Department of Physics, Virginia Tech, Blacksburg, VA 24061, USA}

\author[0000-0001-6421-2449]{Gwen Walker}
\affiliation{Department of Physics, Virginia Tech, Blacksburg, VA 24061, USA}

\author{Luming Sun}
\affiliation{Department of Physics, Anhui Normal University, Wuhu, Anhui 241002, China}

\author[0000-0001-9495-7759]{Lu Shen}
\affiliation{Department of Physics and Astronomy, Texas A\&M University, College Station, TX, 77843-4242 USA}
\affiliation{George P. and Cynthia Woods Mitchell Institute for Fundamental Physics and Astronomy, Texas A\&M University, College Station, TX,10 77843-4242 USA}

\author{Yulong Gao}
\affiliation{School of Astronomy and Space Science, Nanjing University, Nanjing 210093, PR China}
\affiliation{Key Laboratory of Modern Astronomy and Astrophysics (Nanjing University), Ministry of Education, Nanjing 210093, PR China}

\author{Guilin Liu}
\affiliation{CAS Key Laboratory for Research in Galaxies and Cosmology, Department of Astronomy, University of Science and Technology of China, Hefei 230026, China}
\affiliation{School of Astronomy and Space Sciences, University of Science and Technology of China, Hefei 230026, China}

\author{Junfeng Wang}
\affiliation{Department of Astronomy, Xiamen University, Xiamen, Fujian 361005, China}



\begin{abstract}

Measuring the distance of quasar outflows from the central source ($R$) is essential for determining their importance for AGN feedback. There are two methods to measure $R$: 1) A direct determination using spatially resolved Integral Field Spectroscopy (IFS) of the outflow in emission. 2) An indirect method which uses the absorption troughs from ionic excited states. The column density ratio between the excited and resonance states yields the outflow number density. Combined with a knowledge of the outflow’s ionization parameter, $R$ can be determined. Generally, the IFS method probes $R$ range of several kpc or more, while the absorption method usually yields $R$ values of less than 1kpc. There is no inconsistency between the two methods as the determinations come from different objects. Here we report the results of applying both methods to the same quasar outflow, where we derive consistent determinations of $R$ $\approx$ 5 kpc. This is the first time where the indirect absorption $R$ determination is verified by a direct spatially resolved IFS observation. In addition, the velocities (and energetics) from the IFS and absorption data are also found to be consistent. Therefore, these are two manifestations of the same outflow. In this paper we concentrate on the absorption $R$ determination for the outflow seen in quasar 3C 191 using VLT/X-shooter observations.  We also reanalyze an older absorption determination for the outflow based on Keck/HIRES data and find that revised measurement to be consistent with ours. Our companion paper details the IFS analysis of the same object. 
\end{abstract}


\section{Introduction}

    Active galactic nuclei (AGN) feedback serves as an important link in the co-evolution of supermassive black holes (SMBHs) and their host galaxies. It has thus become an essential feature of modern theories of galaxy formation \citep[e.g.,][]{cattaneo2009role, silk2012current, hlavacek2024agn} and state-of-the-art cosmological simulations \citep[e.g.,][]{dubois2012self,dave2019simba,donnari2021quenched}. In the quasar mode of AGN feedback, powerful outflows are driven by SMBHs accreting close to the Eddington limit. The energy and momentum carried by these outflows can be re-deposited in the surrounding medium and thus contribute to various feedback processes. To assess this contribution quantitatively, it is important to determine the spatial extent of the outflowing gas ($R$), its mass  ($M$), mass flow rate ($\dot{M}$) and kinetic luminosity ($\dot{E}_k$) from observations. 
    
    Observational signatures of quasar outflows are detected primarily: (a) in emission, as spatially-resolved galactic scale winds studied using Integral Field Spectroscopy (IFS) (and other techniques such as long-slit spectroscopy) and (b) in absorption, as blueshifted troughs in the quasar spectrum observed as a result of the intersection of our line of sight with the outflowing gas. Both these analysis techniques have been utilized greatly over the last two decades for their distinct advantages. The absorption analysis is much richer in spectroscopic information and has thus allowed the study of a wide range of physical parameters (e.g., density and ionization state) in the outflows. The detection of absorption troughs from metastable excited states plays a crucial role in the analysis as their population ratio with the resonance state is a reliable indicator of the outflow's number density. The ionization state is typically determined through a detailed photoionization modeling of the outflow, and combined with the number density leads (using eqn. \ref{distance_VLT}) to an indirect determination for the distance of the outflow from the central source \citep[e.g.,][]{hamann2001high, de2001keck, moe2009quasar, bautista2010distance, dunn2010quasar, aoki2011outflow,borguet2012major,borguet201210, arav2013quasar, lucy2014tracing, finn2014compact, chamberlain2015large,chamberlain2015strong, miller2018distance,miller2020hst,xu2018mini, xu2020hst, xu2021physical, he2022evidence, byun2022vlt,byun2022vltb, byun2024bal, byun2024extreme,walker2022high, dehghanian2024narrow}. While these studies have revealed a wide range for the distance by placing the outflows at a few pc to several kpc, most of them are still found to be at distances less than a few kpc \citep{arav2018evidence, xu2019vlt}.
    
    IFS analysis on the other hand directly measures the spatial extent of large-scale outflows and provides valuable insights into their morphology and kinematic structure. Recent IFS studies have highlighted the prevalence of kpc scale outflows in AGNs \citep[e.g.,][]{nesvadba2008evidence, alexander2010searching, rupke2011integral, rupke2013multiphase, harrison2012energetic, harrison2014kiloparsec, liu2013observations, liu2014similarity, liu2015integral, karouzos2016unraveling, kakkad2020super, wylezalek2020ionized, kim2023unraveling, shen2023discovery, travascio2024muse, parlanti2024, ubler2024ga}. Emission line diagnostics also exist for determining the electron number density of these outflows \citep{holt2011impact,baron2019discovering}, however they are subject to larger uncertainties as inconsistencies have been revealed between different commonly used methods \citep{davies2020ionized}.

    Although these two methods of distance determinations have led to different scales for the outflows, they are not inconsistent with each other. This is because these methods have either been applied to different objects, or to different outflow phases for the same object \citep[e.g.,][]{noterdaeme2021down, zhao2023discovery, bischetti2024multi}. However, attempts to verify the indirect distance determination from absorption analysis through IFS mapping have also only seen limited success \citep[][also see section \ref{compareothers} for a detailed discussion]{liu2015integral}, thus highlighting the missing link between these two analysis techniques.
    In this paper (and the companion paper by \citealp{zhao3c191}), we present the first consistent distance determination for a quasar outflow from the two analysis techniques, in the form of an ionized outflow in quasar 3C 191. This paper details the absorption analysis for the outflow while its IFS analysis is detailed in \cite{zhao3c191}.

    3C 191 (J2000: RA=08:04:47.97; DEC: +10:15:23.70) is a radio-loud quasar with a bipolar radio structure spanning $\sim$ 5 arcsec ($\sim$ 42 kpc at the quasar redshift) \citep{pearson1985compact, akujor1994images}. It was one of the first quasars to show an extensive absorption system in its spectrum as identified by \citealp{burbidge1966measurement} and \citealp{stockton1966remarkable}. The detection of \ion{Si}{II} ground and excited states with high dispersion observations allowed an estimation for the electron number density with $n_e \sim 10^3$ cm$^{-3}$ and the distance of the outflow component from the central source was determined to be $R \sim$ 10 kpc \citep{bahcall1967analysis, williams1975splitting}. However, \citealp{hamann2001high} analyzed a high-resolution spectrum of 3C 191 and obtained $n_e \sim 300$ cm$^{-3}$ and $R \sim$ 28 kpc for the outflow component. Thus, while there is a slight disagreement between these analysis for the physical parameters (also addressed in section 4.6 of \citealp{hamann2001high}), they nonetheless suggest a galactic scale absorption line outflow. This makes the outflow in 3C 191 a great choice for a complementary analysis using both absorption and IFS techniques.   

    In this paper, we present a study of the X-shooter \citep{vernet2011x} spectrum of the quasar 3C 191 obtained on the Very Large Telescope (VLT). Section \ref{sec:data} describes the observation and how the redshift and normalized spectrum were obtained for our analysis. In Section \ref{sec:spec_analysis}, we analyse the spectrum to extract the column densities of the observed troughs, which allows us to obtain a photoionization model for the outflow and its electron number density. In section \ref{sec:dist_energ}, we determine the distance of the outflow from the central source and its energetics. In section \ref{sec:HamannAnalysis}, we revisit the analysis of \cite{hamann2001high}, and update the outflow parameters obtained by them from the Keck/HIRES spectrum. Section \ref{IFSZhao} compares the results of our absorption analysis of the outflow with the IFS analysis of \cite{zhao3c191}. We discuss important aspects of our results in section \ref{sec:discussion} and conclude the paper with a summary in section \ref{summary}. Throughout the paper, we adopt a standard $\Lambda$CDM cosmology with h = 0.677, $\Omega_m$ = 0.310 and $\Omega_{\Lambda}$ = 0.690 \citep{planck2018}.

\section{VLT X-shooter Analysis} \label{sec:VLT}
   
    \subsection{Data Overview} \label{sec:data}

    \subsubsection{Observation and Data Acquisition}
   
\begin{figure*}
\centering
\includegraphics[width=0.95\linewidth]{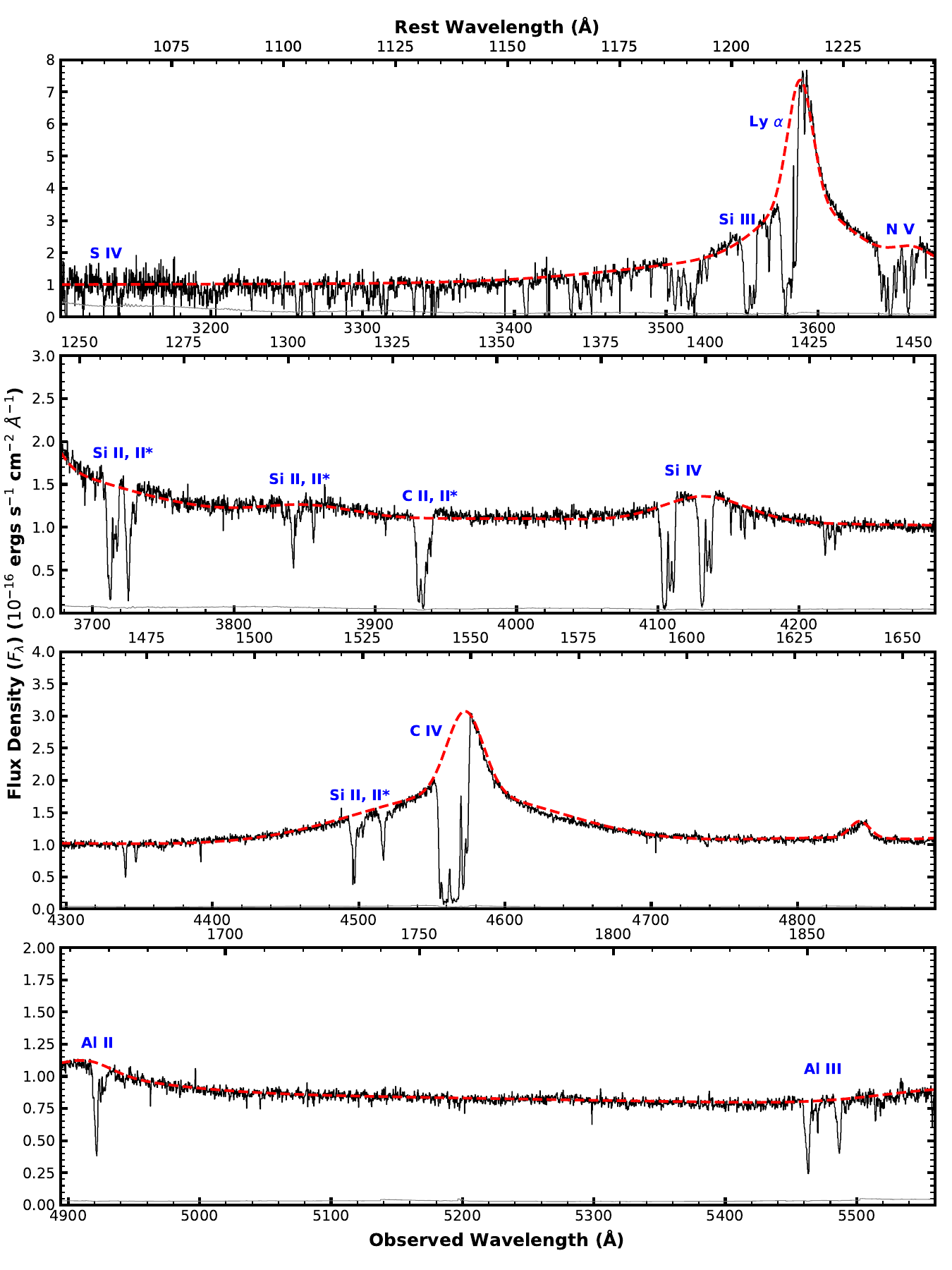}
\caption{VLT/X-shooter spectrum of 3C 191. The red dashed curve shows our modeled continuum along with the emission features. Important absorption features from the identified outflow system are marked in blue. The noise in the VLT/X-shooter flux is shown in gray.\label{fig:spectrum}}
\end{figure*}

   3C 191 was observed with VLT/X-shooter on 03 December, 2013, with a total exposure time of 3000 secs (as part of program 092.B-0393, PI: Leipski). The data covers a spectral range of 2989-10200 \r{A}, with resolution R$\approx$ 4100. 
   The spectrum was processed by ESO through their standard pipelines \citep[see][for a detailed description of the X-shooter pipeline]{modigliani2010x} to remove instrument and atmospheric signatures and perform flux and wavelength calibration. The processed spectrum was then made available on the ESO Science archive\footnote{\url{https://archive.eso.org/scienceportal}}. Several noise spikes less than 3 pixels wide (originating from impact of cosmic rays with the detector) were seen throughout the processed spectrum. We omitted these affected pixels using three sigma clipping and obtained the finalized spectrum (shown in Fig. \ref{fig:spectrum}) for our analysis.
   
   \subsubsection{Redshift Determination}
   \cite{zhao3c191} determined the redshift of 3C 191 using the narrowest component of the [\ion{O}{III}] 5007 \r{A} emission line extracted from the central 0.25" around the quasar nucleus. They determined the systematic redshift of the quasar to be $z$ = 1.9527 $\pm$ 0.0003 (see their section 3.1 for a detailed description of the emission model). This was also found to be in agreement with the redshift determined from [\ion{O}{II}] $\lambda \lambda$ 3727, 3729 doublet ($z$ = 1.9527 $\pm$ 0.0003) detected in emission in the VLT/X-shooter spectrum and therefore we adopt this redshift for our analysis.

   \subsubsection{Unabsorbed Emission Modeling}
   In order to identify the troughs and determine their column densities, we need to obtain a normalized spectrum by dividing the data by the unabsorbed emission model. In the spectrum of 3C 191, we first select spectral regions free from significant emission or absorption ($\lambda_{obs}$ $\simeq$ 3165, 3370, 4000, 4300 and 5300 \r{A}) to use as anchor points for our continuum fit. We find that a single power law is unable to describe the emission in the observed wavelength range analysed in this paper and therefore we employ a broken power law of the form \(F_{\lambda} = F_{4000} (\lambda_{obs}/4000)^{\alpha}\), where $\alpha$ takes different values before ($\alpha_1$) and after ($\alpha_2$) $\lambda_{obs}$ = 4000 \r{A}. The best fit parameters for the model obtained using non-linear least squares are: $\alpha_1$ = 0.34$\pm$0.09, $\alpha_2$ = $-$1.08$\pm$0.01 and $F_{4000}$ = 1.104 ($\pm$ 0.003) $\times 10^{-16}$ ergs s$^{-1}$ cm$^{-2}$ $\r{A}^{-1}$ with $\chi_{red}^2$ (reduced chi-square) = 0.754. \par
    The X-Shooter spectrum also shows clear signature of many prominent emission features and thus we obtain a best-fit model using non-linear least squares for each of these features by considering absorption free regions in their vicinity. The \ion{Ly}{$\alpha$}/\ion{N}{V} complex is modeled with four independent Gaussian components. The \ion{C}{IV} emission feature also requires separate broad emission line (BEL) and narrow emission line (NEL) components and was therefore modeled using two independent Gaussians. The \ion{Si}{IV} emission feature on the other hand, is well modeled with a single Gaussian component. Finally, we also detect weak emission in two regions: 3800 \r{A} $\lesssim$ $\lambda_{obs}$ $\lesssim$ 3900 \r{A} and 4700 \r{A} $\lesssim$ $\lambda_{obs}$ $\lesssim$ 5000 \r{A}. Based on the composite quasar SDSS spectra of \cite{berk2001composite}, the first region is likely to be a blend of \ion{O}{I}/\ion{Si}{II}, whereas the latter could contain contribution from \ion{He}{II} along with a blend of \ion{O}{III}]/\ion{Al}{II}/\ion{Fe}{II}. We are able to model these regions with one and three independent Gaussians respectively. Table. \ref{table:emfit} reports the wavelength ranges (rounded off to the nearest integer) used as anchor for fitting each emission complex and the $\chi_{red}^2$ for the respective fits. 
    Combining the continuum model and the different emission components gives us the unabsorbed emission model, which is shown in Fig. \ref{fig:spectrum} by the red dashed curve. 

    \begingroup
\renewcommand{\arraystretch}{1.15}
\begin{table}
\caption{Wavelength ranges used for fitting the emission lines in the VLT/X-Shooter spectrum and the resulting $\chi_{red}^2$. \label{table:emfit}}
\centering
\begin{tabular}{|c|c|c|c|}
\hline
Complex & Lines & Wavelength Range (\r{A}) & $\chi_{red}^2$ \\
\hline
\multirow{4}{*}{1}& \ion{Ly}{$\alpha$} + \ion{N}{V} &3480-3485, 3493-3498 &1.261 \\
&  &3533-3543, 3562-3565&  \\
& & 3595-3636, 3667-3700   & \\
& & 3740-3770& \\
\hline
2 & \ion{O}{I} + \ion{Si}{II} & 3795-3835, 3870-3905 &0.656 \\
\hline
\multirow{2}{*}{3}& \ion{Si}{IV} &4075-4090, 4117-4123&0.691 \\
& &4140-4146, 4180-4200 & \\
\hline
\multirow{2}{*}{4}& \ion{C}{IV} & 4400-4480, 4530-4540  &0.944 \\
& &4580-4630 & \\
\hline
\multirow{2}{*}{5}&\ion{He}{II} + \ion{O}{III}]&4672-4732,4745-4880 & 0.999 \\
&+ \ion{Al}{II} + \ion{Fe}{II}&4900-4910, 4931-4990 & \\

\hline
\end{tabular}
\end{table}
\endgroup

   \subsection{Spectral Analysis} \label{sec:spec_analysis}
   
   \subsubsection{Identifying Outflow Systems}

   Having determined the systematic redshift and modeled the unabsorbed emission, we identify the main absorption system for the low ionization species (\ion{Si}{ii}, \ion{C}{ii}, \ion{Al}{ii}, \ion{AL}{iii} and \ion{Si}{iii}), spanning a velocity range $-$950 $\lesssim$ $v$ $\lesssim$ $-$500 km s$^{-1}$, with the deepest absorption at  $v \sim -720$ km s$^{-1}$. The same system is detected in the high ionization species (\ion{C}{iv}, \ion{Si}{iv}, \ion{S}{iv} and \ion{N}{v}) as well. However, the high ionization troughs are broader and show a shift in velocity space with the deepest absorption at  $v \sim -800$ km s$^{-1}$. Similar shift in velocities for higher ionization potential ions has been previously reported in absorption troughs from an outflow in PKS J0352-0711 by \cite{miller2020hst}. Based on the width of the \ion{C}{IV} trough ($\Delta v$ $\sim$ 1500 km s$^{-1}$), the outflow system is characterized as a mini broad absorption line \citep[mini-BAL;][]{hamann2004asp}. 

   \subsubsection{Column Density Determination} \label{sec:colden}

   The ionic column densities $N_{ion}$ of the various observed species can be obtained from the troughs by assuming an absorber model for the cloud.  The simplest model, known as the apparent optical depth (AOD) model assumes a homogeneous outflow that completely covers the source. In this case, the normalized intensity profile as a function of velocity $I(v)$ is related to the optical depth $\tau(v)$ as $I(v) = e^{-\tau(v)}$. $N_{ion}$ for a transition with rest wavelength $\lambda$ (in Angstroms) and oscillator strength $f$ is then given as \citep{savage1991analysis}:
    \begin{equation}{\label{nion}}
    N_{ion} = \frac{m_e c}{\pi e^2 f \lambda} \int \tau(v) \textrm{ dv} = \frac{3.8 \times 10^{14}}{f \lambda} \int \tau(v) \textrm{ dv} \textrm{\hspace{0.3cm}[cm$^{-2}$]}
    \end{equation}
    where $m_e$ is electron mass, $c$ is the speed of light, $e$ is elementary charge. The homogeneous AOD model does not account for partial line of sight covering of the outflow and line saturation and therefore can only lead to lower limits for $N_{ion}$ in most cases. If troughs from two or more lines corresponding to the same lower energy level of an ion are observed, we can employ a partial covering (PC) model. In this two-parameter model, the absorber is assumed to be covering a (velocity dependent) fraction $C(v)$ of a constant emission source. The normalized intensity for the two lines are then given as:
    \begin{equation} \label{I1}
    \begin{aligned} 
    I_1(v) = 1 - C(v) + C(v)\cdot e^{-\tau(v)} \\
    I_2(v) = 1 - C(v) + C(v)\cdot e^{-R \tau(v)}
    \end{aligned}    
    \end{equation}
    where $R$ is the expected line strength ratio given by $R = (g_i f_{ik} \lambda)_2/(g_i f_{ik} \lambda)_1$ (where $g_i$ is the degeneracy of the lower level $i$, $f_{ik}$ is the oscillator strength of the transition between levels $i$ and $k$, and $\lambda$ is the transition's wavelength). Having obtained $I_1(v)$ and $I_2(v)$ from the spectrum, $C(v)$ and $\tau(v)$ can be obtained numerically by solving for eqns. (\ref{I1}) simultaneously \citep[see][for a detailed desciption of the model]{arav1999hubble,arav2005x}. \par
\begin{figure}
\centering
\resizebox{\hsize}{!}
{\includegraphics[width=1.20\linewidth]{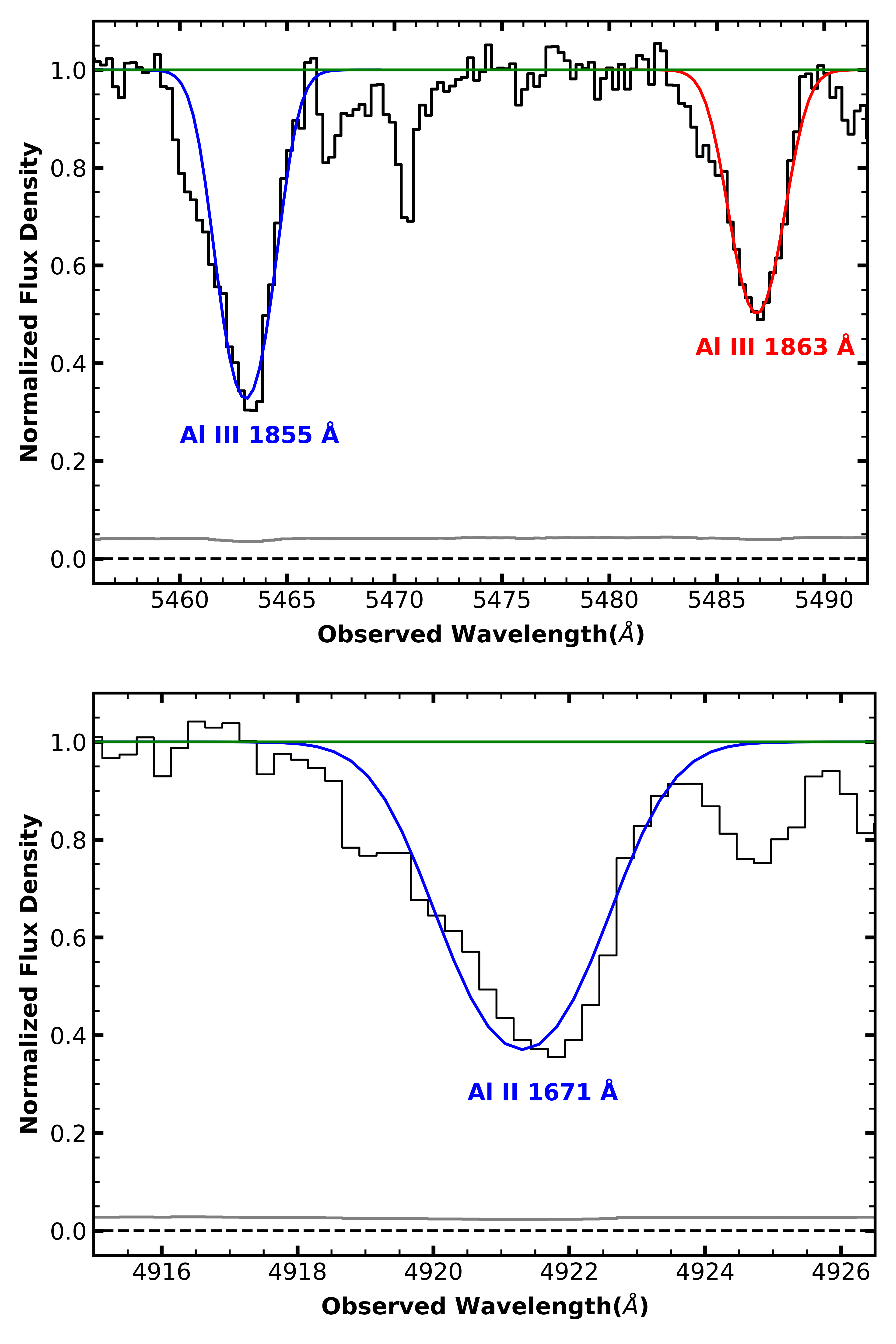}}
\caption{Detected troughs for the \ion{Al}{III} 1855 and 1863 \r{A} (top) and \ion{Al}{II} 1671 \r{A} (bottom) transitions along with their Gaussian models. The Gaussian models are based on the \ion{Al}{III} 1863 \r{A} trough, and then its template with fixed centroid and width is scaled to match the depth of the \ion{Al}{III} 1855 \r{A} and the \ion{Al}{II} 1671 \r{A} trough. The solid green lines represents the local continuum model. The gray line shows the noise in the VLT/X-shooter flux. \label{fig:Al}}
\end{figure}

    We model the troughs in velocity space using a Gaussian profile for the optical depth and note that for most of the observed ionic species (\ion{H}{I}, \ion{N}{V}, \ion{C}{II}, \ion{C}{IV}, \ion{Si}{III} and \ion{Si}{IV}), the troughs appear saturated and therefore only lower limits for $N_{ion}$ can be obtained for them based on the AOD model. We detect troughs from two different ionized species of \ion{Al}{}: \ion{Al}{iii} $\lambda \lambda$ 1855, 1863 and \ion{Al}{ii} 1671 \r{A}. We first model the \ion{Al}{iii} 1863 \r{A} trough with a Gaussian profile and then use this model as a template for the \ion{Al}{III} 1855 \r{A} and \ion{Al}{II} 1671 \r{A} troughs by fixing the centroid and width of the profile while allowing its depth to vary independently (see Appendix \ref{appendix} for an alternate multi-component modeling and the difference between the two approaches). The best-fit models (shown in Fig. \ref{fig:Al}) are determined by non-linear least squares. We note that for \ion{Al}{II}, the deepest part of the model and the trough show a slight offset ($\sim$ 30 km s$^{-1}$). However, as this shift is less than one resolution element in the velocity space ($\sim$ 73 km s$^{-1}$), we continue to use this model for our analysis as it is physically motivated. As the two \ion{Al}{III} lines originate from the same lower energy level, we can use the partial covering method to obtain a measurement for its $N_{ion}$. We find that the $N_{ion}$ determined using the PC method differs by less than 20 \% from the AOD determination, thus indicating that these troughs are not affected greatly by non-black saturation. As the \ion{Al}{ii} 1671 \r{A} trough appears shallower than the \ion{Al}{iii} 1855 \r{A} trough, it would be affected even less by non-black saturation \citep[see section 2.2 in][for a detailed description of non-black saturation]{arav2018evidence}. This allows us to use the $N_{ion}$ determined from the AOD method as a measurement. Reliable measurements of column densities from two different ionization states of Al plays an important role in our photoionization analysis, which we describe in detail in the next section.
    
\begin{figure}
\centering
\resizebox{\hsize}{!}
{\includegraphics[width=1.00\linewidth]{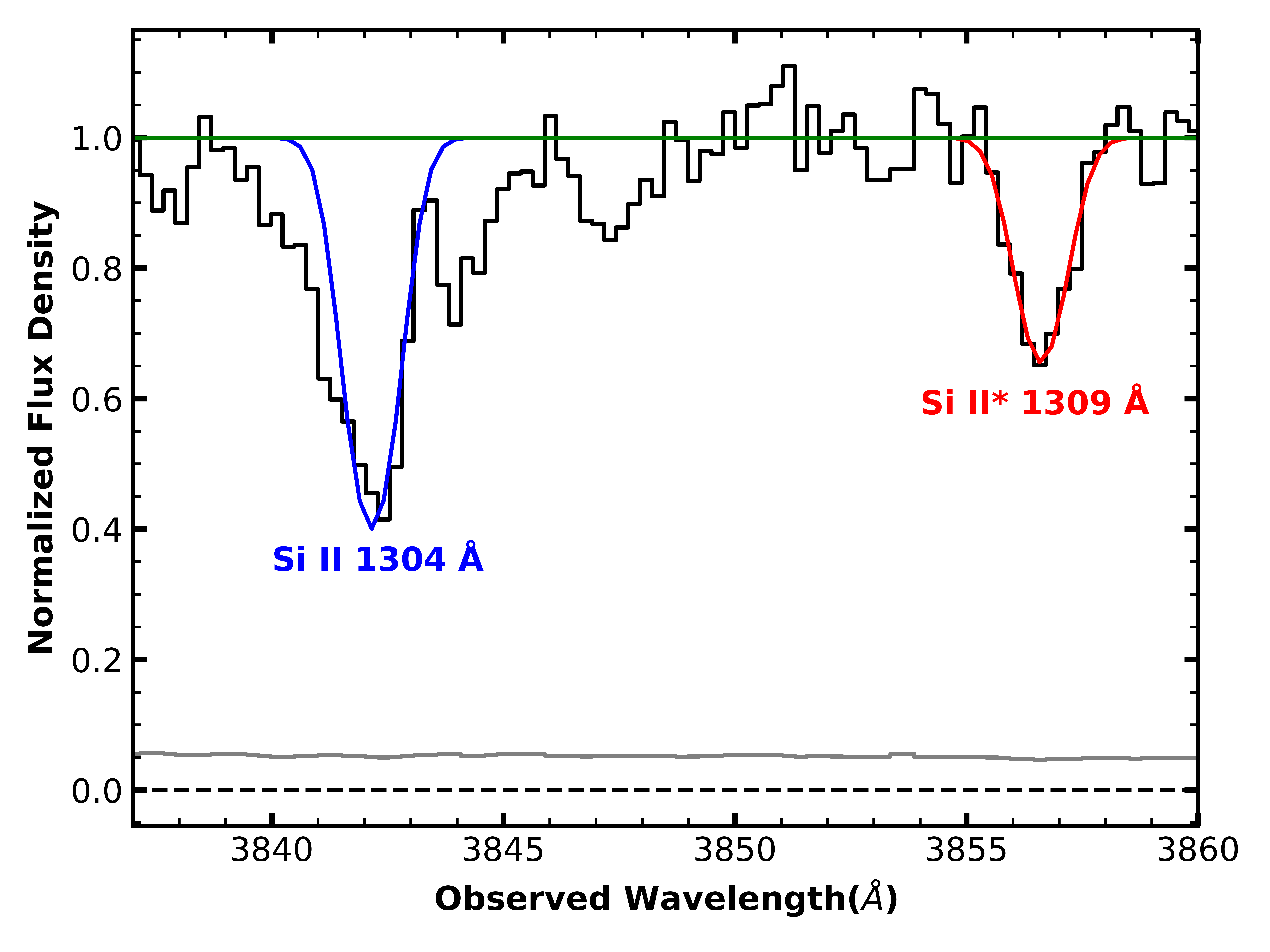}}
\caption{Observed troughs for the \ion{Si}{II} 1304 \r{A} and \ion{Si}{II*} 1309 \r{A} transitions along with their Gaussian models. The model is based on the \ion{Si}{II*} excited trough, and then its template with fixed centroid and width is scaled to match the depth of the \ion{Si}{II} trough. The solid green line represents the local continuum model. The gray line shows the noise in the VLT/X-shooter flux. \label{fig:siII}}
\end{figure}

   For \ion{Si}{II}, we detect multiple troughs originating from both the ground state and the 287 cm$^{-1}$ metastable excited state. Among the detected \ion{Si}{II} transitions, the 1304 \r{A} line has the smallest oscillator strength. It also appears much shallower than the other deeper \ion{Si}{II} troughs (at 1260 and 1527 \r{A}) and can thus be considered unsaturated. Using the same argument to compare the different \ion{Si}{II}* transitions at 1265, 1309 and 1533 \r{A} shows the 1309 \r{A} trough can also be considered as unsaturated. Therefore, the AOD column densities derived from the \ion{Si}{ii/ii*} 1304/1309 \r{A} transitions can be used as measurements. Our modeling of these troughs follows the same procedure as that of the \ion{Al}{} troughs. We first model the \ion{Si}{II}* 1309 \r{A} trough with a Gaussian profile and use this as the template for obtaining the best-fit model for the  \ion{Si}{II} 1304 \r{A} trough (shown in Fig. \ref{fig:siII}). 
    
    We summarize the determined column density measurements and limits for the detected troughs in Table. \ref{table:colden}, along with the predictions of our best-fit photoionization model (section \ref{photoion}). The reported errors include the uncertainty in determining the best-fit model for the unabsorbed emission. We first obtain the 1-$\sigma$ errors for the different components of our best-fit emission model (shown in red in Fig. \ref{fig:spectrum}). Using them we construct two additional continuum models in which each component is shifted by + $\sigma$ and $-\sigma$, respectively. We use these new continuum levels to re-normalize the VLT/X-Shooter spectrum individually and obtain the column densities of the different ionic species for them. Their deviation from the column densities obtained for the best-fit emission model (reported in Table. \ref{table:colden}) allows us to estimate their uncertainty due to our modeling of the emission features. We add them in quadrature with the 1-$\sigma$ uncertainties obtained similarly for our modeling of the absorption troughs. For \ion{Al}{II}, \ion{Al}{III} and \ion{Si}{II}, we also include the uncertainty due to our choice of models for their troughs as outlined in Table. \ref{table:model_comp} (see Appendix \ref{appendix} for a comparison between the single component and multi-component approach).

\begingroup
\setlength{\tabcolsep}{10pt}
\renewcommand{\arraystretch}{1.6}
\begin{table}
\caption{Ionic column densities for the outflow in 3C 191. \label{table:colden}}
\centering
\begin{tabular}{lcc}
\hline
Ion        & log($N_{ion}$)$^{a}$ [cm$^{-2}$] & log($N_{mod}$)$^{b}$ [cm$^{-2}$]\\
\hline
\ion{H}{i} & $>14.93^{}_{-0.10}$ & 17.28 \\
\ion{C}{ii} & $>14.90^{}_{-0.09}$ & 15.21\\
\ion{C}{ii*} & $>15.07^{}_{-0.08}$ & - \\
\ion{C}{iv} & $>15.53^{}_{-0.10}$ & 16.30 \\
\ion{N}{v} & $>15.31^{}_{-0.08}$ & 14.63 \\
\ion{Al}{ii} & $13.29^{+0.06}_{-0.07}$ & 13.29\\
\ion{Al}{iii} & $13.96^{+0.07}_{-0.09}$ & 13.96\\
\ion{Si}{ii} & $14.48^{+0.04}_{-0.04}$ & 14.25\\
\ion{Si}{ii*} & $14.21^{+0.05}_{-0.07}$ & - \\
\ion{Si}{iii} & $>14.26^{}_{-0.08}$ & 15.46 \\
\ion{Si}{iv} & $>14.81^{}_{-0.07}$ & 15.56\\
\ion{S}{iv} & $14.89^{+0.08}_{-0.10}$ & 15.29\\
\ion{S}{iv*} & $<13.89^{+0.09}_{}$ & -\\
\hline
\end{tabular}
\tablecomments{($a$) Measured column densities for the outflow from the VLT/X-shooter spectrum. ($b$) Ionic column densities predicted by the best-fit \textsc{Cloudy} model shown by the black dot in Fig. \ref{fig:nvu}. The model values for \ion{C}{ii}, \ion{Si}{ii} and \ion{S}{iv} include the contribution from their excited states.}
\end{table}
\endgroup

\subsubsection{Photoionization modeling} \label{photoion}
    
\begin{figure}
   \centering
   \resizebox{\hsize}{!}
    {\includegraphics[width=1.20\linewidth]{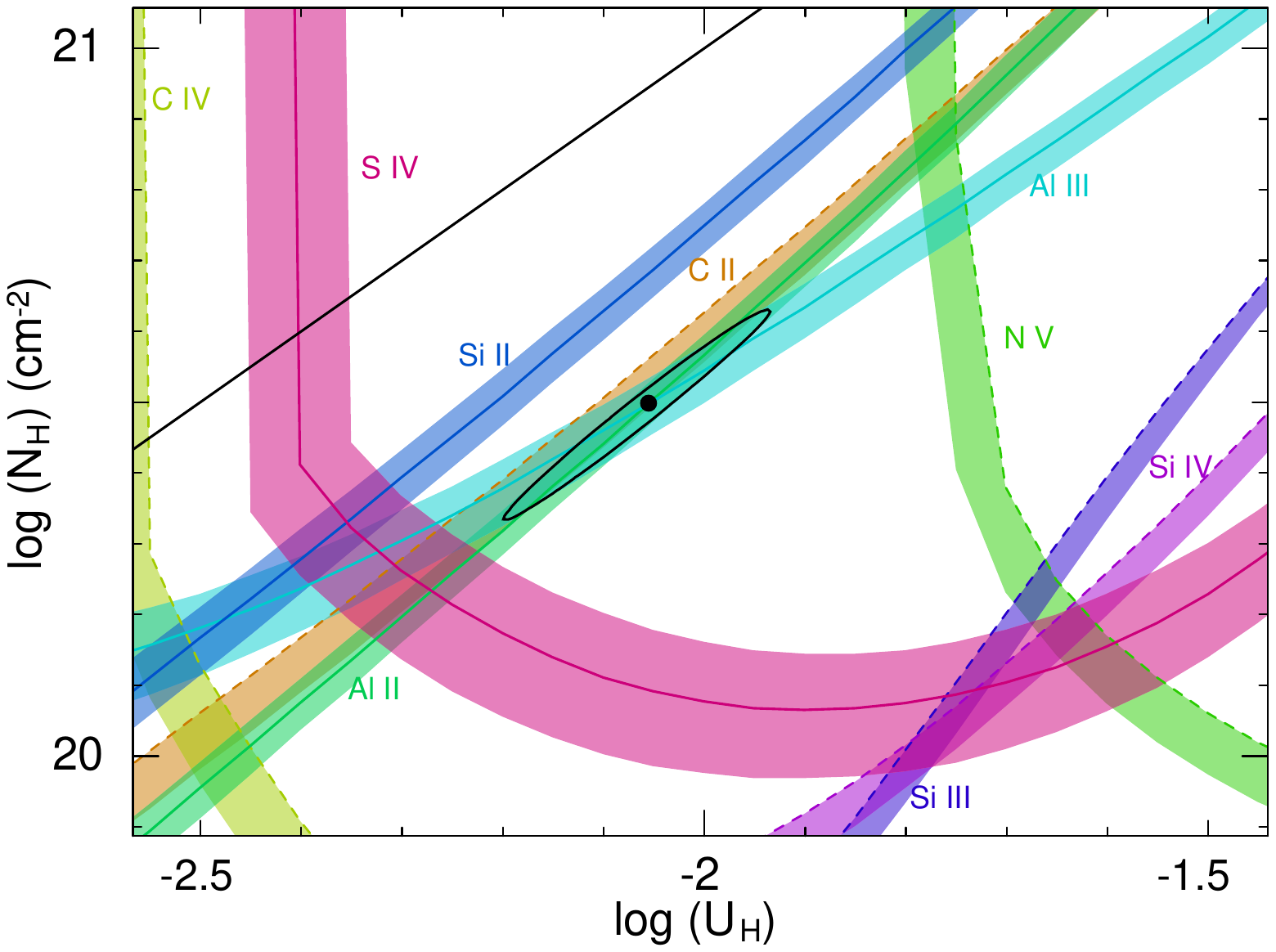}}
      \caption{Plot of Hydrogen column density ($\log{N_H}$) vs. ionization parameter ($\log{U_H}$), with constraints based on the measured ionic column densities. Measurements are shown as solid curves, while the dashed curves show lower limits which allow the parameter space above them. The shaded regions denote the errors associated with each $N_{ion}$ as reported in Table \ref{table:colden}. The phase space solution with minimized $\chi^2$ is shown as a black dot surrounded by a black eclipse indicating the 1-$\sigma$ error. This solution is based on the $N_{ion}$ of \ion{Al}{II} and \ion{Al}{III} (see section \ref{photoion}). The black line marks the position of the Hydrogen ionization front. \label{fig:nvu}}
   \end{figure}

    Ionized outflows are dominated by photoionization equilibrium \citep[and references therein]{davidson1979emission, krolik1999active, osterbrock2006astrophysics} and are thus characterized by their total hydrogen column density ($N_H$) and ionization parameter ($U_H$), which is related to the rate of ionizing photons emitted by the source ($Q_H$) by:
    \begin{equation} \label{uh}
    U_H = \frac{Q_H}{4\pi R^2 n_H c}
    \end{equation} 
    where $R$ is the distance between the central emission source and the observed outflow component, $c$ is the speed of light, and $n_H$ is the hydrogen number density. $Q_H$ is determined by: (a) the choice of the spectral emission distribution (SED) that is incident upon the outflow, (b) the observed flux of the object's continuum at a specified wavelength, (c) the redshift of the object and (d) the choice of cosmology. Having determined the column densities of the observed ionic species, we can constrain the physical state of the gas, using the spectral synthesis code \textsc{Cloudy} \citep[vers. C23.01][]{cloudy2023} which solves the equations of photoionization equilibrium in the outflow. The outflow is modeled as a plane parallel slab with a constant $n_H$ and solar abundance, and is irradiated upon by the modeled SED from the quasar HE0238-1904 \citep{arav2013quasar}, which is the best empirically determined SED in the extreme UV where most of the ionizing photons come from. Fig. \ref{fig:nvu} shows the result of our photoionization modeling.
    Using the methodology described by \cite{edmonds2011galactic}, $N_H$ and $U_H$ are varied in steps of 0.1 dex, keeping all other parameters ($n_H$, abundances and the incident SED) constant, leading to a two dimensional grid of models in the parameter space with predictions for all the $N_{ions}$ in the slab. The different colored contours represent the allowed values in the $(N_H, U_H)$ phase space which correspond to the observed column density constraints for the ionic species. The strongest such constraint comes from the column density measurements of the two Al ions. The ratio of the column densities of \ion{Al}{ii} and \ion{Al}{iii} serves as an indicator of the ionization state of the outflow which is independent of the assumed abundances. Therefore based only on the  \ion{Al}{ii} and \ion{Al}{iii} $N_{ion}$ measurements, we obtain log $N_H$ = $20.50^{+0.13}_{-0.16}$ [cm$^{-2}$] and log $U_H$ = $-2.06^{+0.12}_{-0.14}$. This solution is shown in Fig. \ref{fig:nvu}, surrounded by an error ellipse that contains the phase space values within one standard deviation of our best-fit solution. We use this solution for the rest of our analysis. As can be seen, this solution underpredicts the \ion{Si}{II} column density (traced by the blue curve). Furthermore, the constraint for \ion{Si}{ii} is almost parallel to that of \ion{Al}{ii} and therefore they cannot both simultaneously be satisfied by any point in the ($N_H, U_H$) parameter space. The simplest way to resolve this conflict is to postulate an abundance ratio of Si to Al that differs from the one observed in the Sun. We find that an increase of $\sim$ 0.4 dex in the relative abundance Si/Al (with respect to solar abundances) would ensure that the measured column density for \ion{Si}{ii} matches our prediction from the model solution. Similarly, our model overpredicts the \ion{S}{IV} column density (traced by the pink curve). However, the measurement can be brought in complete agreement with our model with a decrease of $\sim$ 0.35 dex in the relative abundance of S with respect to Al. The proposed changes for both Si/Al and S/Al abundance ratios are well within the ranges allowed by empirical abundance models in quasar outflows \citep{arav2013quasar}. Finally, Fig. \ref{fig:nvu} shows that our model solution satisfies the lower limit constraints (represented by dashed curves) from all ionic species except \ion{N}{V}. We note that \ion{N}{V} has an ionization potential (IP) of $\sim$ 98 eV and is thus associated with a region with higher ionization than the other observed ionic species in our analysis which have IP $\lesssim$ 64 eV.  Observations of quasar outflows in the extreme UV have shown that when reliable column densities for higher ionization species (e.g., \ion{Ne}{VIII} and \ion{Mg}{X}) can be obtained, a single ionization phase is unable to provide an acceptable physical model for the outflow, thus requiring two distinct ionization phases \citep{arav2013quasar, arav2020hst}. The high ionization species then correspond to a phase with higher $N_H$ and $U_H$. From the VLT/X-shooter spectrum of 3C 191, we cannot obtain reliable measurements for column densities of other high ionization species, and therefore cannot constrain the high ionization phase completely. However, \ion{N}{V} can still be ascribed to this phase, and that would explain why its $N_{ion}$ constraint is not satisfied by our solution which represents the low ionization phase.

   \subsubsection{Electron Number Density} \label{sec:eden}
    The ratios of the column densities of excited to resonance lines are an indicator of the electron number density ($n_e$) of an outflow under the assumption of collisional excitation \citep[e.g.,][]{arav2018evidence}. The identified outflow system in 3C 191 contains multiple troughs corresponding to the ground and excited states of \ion{Si}{II}. As discussed in section \ref{sec:colden}, the \ion{Si}{ii} 1304 \r{A} and \ion{Si}{ii*}  1309 \r{A}  transitions yield the most reliable measure of the column densities of \ion{Si}{ii} and \ion{Si}{ii*}. We obtain theoretical population ratios between the ground and excited state as a function of $n_e$ using the \textsc{Chianti} atomic database \citep[vers. 9.0][]{dere1997chianti,dere2019chianti}.

\begin{figure}
\centering
\resizebox{\hsize}{!}
{\includegraphics[width=1.00\linewidth]{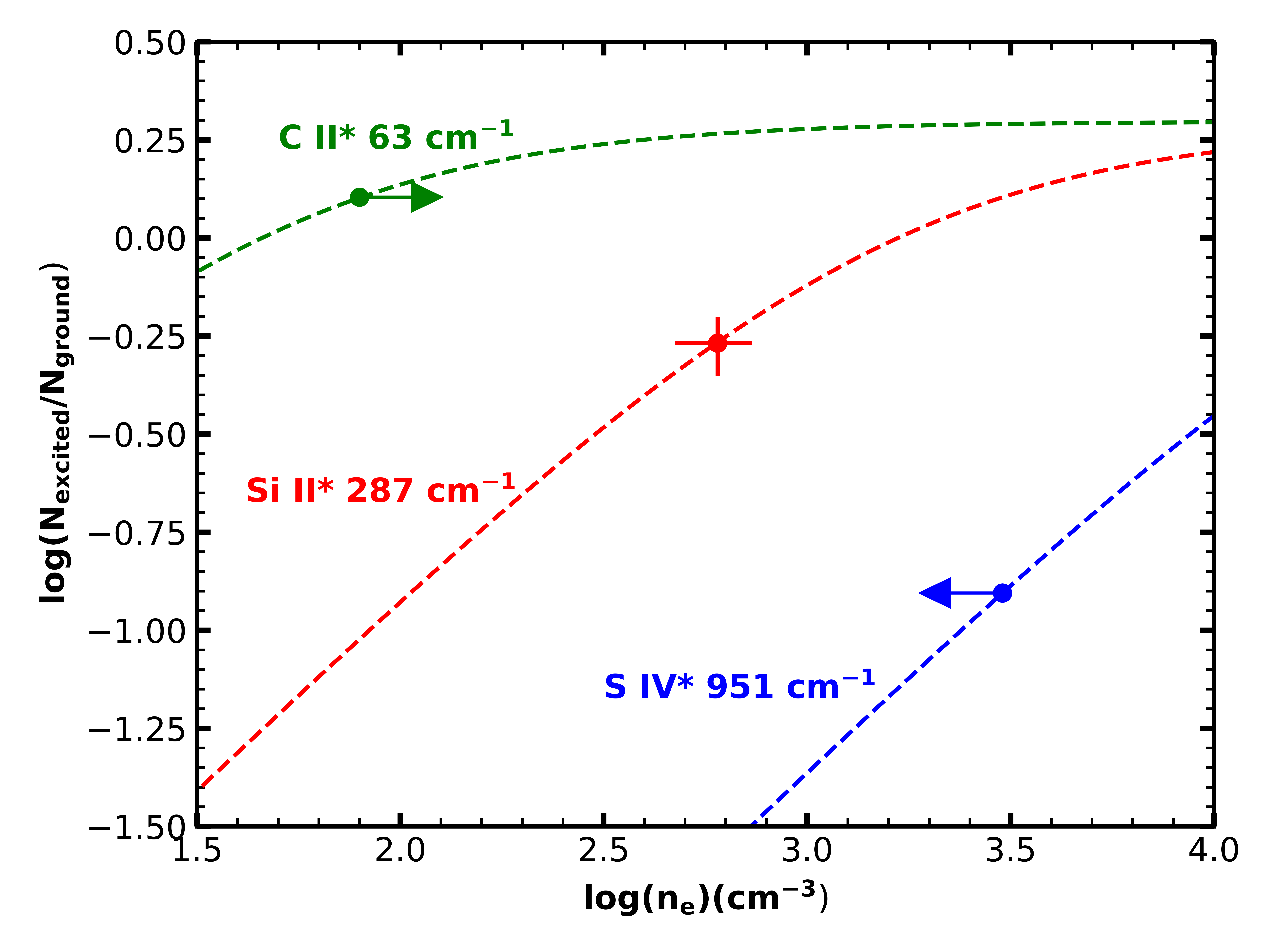}}
\caption{Dashed curves represent the theoretical population ratios for the transitions for an effective electron temperature $T_e$ = 8500 K (determined from our \textsc{Cloudy} modeling). The solid horizontal lines represent the observed column density ratios and their corresponding errors/limits. \label{fig:Chianti}}
\end{figure}

   A comparison of these theoretical predictions with the observed ratio of column densities of \ion{Si}{ii*} and \ion{Si}{ii} is shown in Fig. \ref{fig:Chianti}. This leads to log($n_e$) = $2.78^{+0.08}_{-0.10}$ [cm$^{-3}$]. We also detect \ion{C}{ii} 1335 \r{A} and \ion{C}{ii*} 1336 \r{A} transitions, which appear to be saturated and thus their AOD column densities will correspond to lower limits. Their ratio provides a lower limit of log($n_e$)  $\geq 2.12^{}_{-0.22}$ [cm$^{-3}$]. Although we do not detect the \ion{Si}{iv*} 1073 \r{A} trough, we use the Gaussian template obtained from the \ion{Si}{iv} 1063 \r{A} trough to obtain an upper limit for the column density of \ion{Si}{iv*} based on its non-detection in the spectrum. The ratio of the \ion{Si}{iv*} and \ion{Si}{iv} column densities can then be used to determine an upper limit of log($n_e$) $\leq 3.38^{+0.10}_{}$ [cm$^{-3}$]. Fig. \ref{fig:Chianti} shows that the obtained upper and lower limits are in agreement with our $n_e$ determined from \ion{Si}{II*} and \ion{Si}{ii} population ratio. 

    \subsection{Distance and Energetics} \label{sec:dist_energ}

   \subsubsection{Distance Determination} \label{sec:distance}

   Having determined $U_H$ and $n_e$, we can use eqn. (\ref{uh}) to obtain the distance between the outflow and the central source ($R$) as:

   \begin{equation} \label{distance_VLT}
    R = \sqrt{\frac{Q_H}{4\pi U_H n_H c}}
    \end{equation}

    $Q_H$ can be obtained by integrating over the SED for energies above the Rydberg limit:

    \begin{equation} \label{Qh}
        Q_H = \int^{\infty}_{\nu_{Ryd}} \frac{L_{\nu}}{h \nu} d\nu
    \end{equation}

\begin{figure}[h]
   \centering
   \resizebox{\hsize}{!}
        {\includegraphics[width=1.00\linewidth]{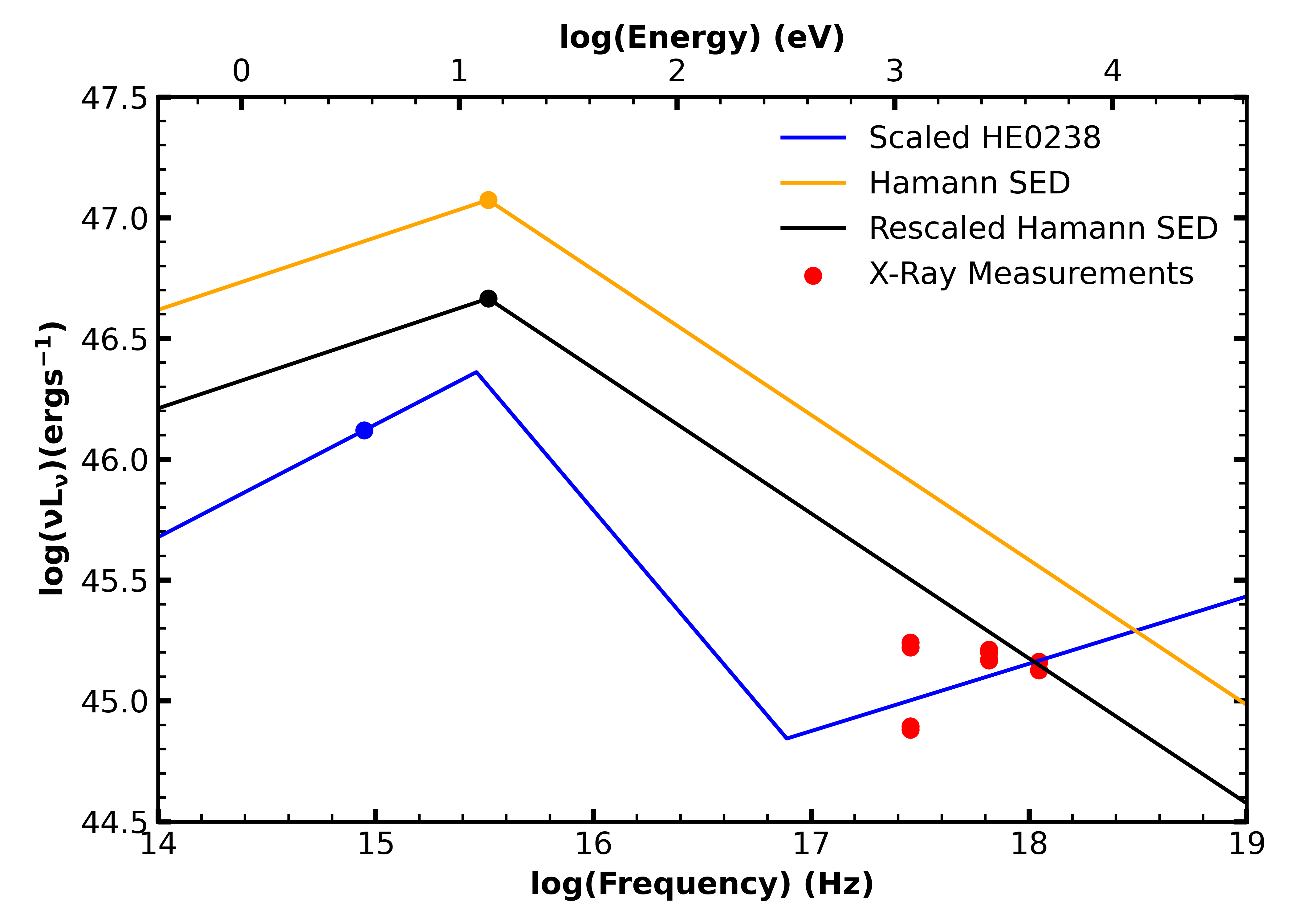}}
      \caption{SEDs considered in the analysis. The blue curve corresponds to the scaled HE0238 SED, with the VLT/X-shooter flux used for scaling (at $\lambda_{obs}$ = 10000 \r{A}) given by the blue dot. The orange curve shows the SED used by \cite{hamann2001high} and the black curve shows the Hamann SED rescaled based on the VLT/X-shooter flux (see section \ref{HamannSED} for more details). The red dots show the available X-ray flux measurements for the quasar. The error on these measurements are between 5\% and 15\% and thus comparable to or smaller than size of the points. \label{fig:SED}}
   \end{figure}

    The spectrum obtained from quasars with low-ionization absorption line outflows can be strongly affected by extinction due to dust in the host galaxy \citep{hall2002unusual, dunn2010quasar, leighly2024physical}. To account for this reddening, we follow the methodology described by \cite{hall2002unusual}, which includes dereddening the spectrum until the continuum slope matches that of the  composite SDSS quasar of \cite{berk2001composite}. In doing so, we find that the slope of the composite SDSS quasar matches that of the HE0238 SED for frequencies smaller than that of the Ly $\alpha$ line. The HE0238 SED is thus representative of a typical dereddened quasar spectrum. Therefore, to obtain the dereddened SED incident on the cloud, we scaled the HE0238 SED, by matching its flux to the observed flux from the VLT/X-shooter spectrum at an observed wavelength of $\lambda_{obs}$ = 10,000 \r{A} ($\lambda_{rest}$ = 3386.78 \r{A}). We chose the longest available wavelength for scaling as AGN extinction curves show that the extinction generally increases with increasing wavenumber, and is therefore smaller for longer wavelengths \citep{czerny2004extinction, gaskell2004nuclear}. The resulting scaled SED is shown in blue in Fig. \ref{fig:SED}. 
    The ionizing photons come from wavelengths shorter than the Lyman Limit which is not covered by VLT/X-shooter. Therefore to get a handle on the ionizing continuum we use the available \textsc{Chandra} observations of the object covering the 0.2-2 keV range, obtained from the second release of the Chandra Source Catalog (CSC).\citep{evans2010chandra, 2020AAS...23515405E} The photometric flux measurements (converted to luminosity) are shown in the quasar rest frame in Fig. \ref{fig:SED} as red points. The X-ray measurements are in agreement with the prediction from the scaled HE0238 SED, further strengthening our case for its use to model the incident flux on the cloud. Therefore, based on our scaled SED, we obtain $Q_H$ = $4.43_{-0.07}^{+0.07}$ $\times$ $10^{56}$ $s^{-1}$ and Bolometric Luminosity $L_{Bol}$ = $8.00_{-0.13}^{+0.13}$ $\times$ $10^{46}$ erg $s^{-1}$. 

    Typically, $n_H$ is estimated from $n_e$ under the assumption of highly ionized plasma, which leads to the approximation $n_e\approx 1.2 n_H$. However, it has been shown that for the low ionization phases of the outflow, the formation of the hydrogen ionization front can lead to a sudden drop in $n_e$ and therefore an underestimation for $n_H$ \citep{Sharma2024a}. We thus run a \textsc{Cloudy} model for our outflow with log($n_H$) = log($n_e$) = 2.78 and the parameters obtained in section \ref{photoion} and find that for this cloud, the hydrogen ionization front is not entirely developed, and thus while $n_e$ deviates slightly from assumption of highly ionized plasma, the drop is not significant and an average over the cloud based on the zone-wise \ion{Si}{ii} column density yields $n_e\approx$ 1.1 $n_H$. Substituting the determined values of $Q_H, U_H$ and $n_H$ into eqn. \ref{distance_VLT}, we find that the outflow is located $5.1^{+0.9}_{-1.0}$ kpc away from the central source. 

    To obtain an estimate of the systematic error in the distance determination due to different SEDs, we check the difference between the $Q_H$ and $U_H$ determined from the HE0238 SED and the SED used by \cite{hamann2001high} in their Keck/HIRES analysis. The latter leads to a 25\% increase in $Q_H$ and a $7\%$ decrease in $U_H$. Therefore, using eqn. (\ref{distance_VLT}), we can estimate the systematic error in our distance determination to be $\sim$ 16\%. 

    \subsubsection{Energetics}

    The outflow can be modeled as a partial thin spherical shell covering a solid angle $\Omega$, and moving with velocity $v$ \citep{borguet201210}. The total mass of the gas within the outflow can then be obtained as:

    \begin{equation}
    M \simeq 4\pi\Omega R^2 N_H \mu m_p    
    \end{equation}

    where $R$ is the spatial extent of the outflow, $N_H$ is the total hydrogen column density, $m_p$ is the proton mass, and $\mu$ = 1.4 is the atomic weight of the plasma per proton. Defining the dynamic time scale as the time it takes the gas from the nucleus travelling at an average velocity $v$ to reach the location of the outflow as $t_{dyn}$ = $R/v$, we can obtain the mass loss rate $\dot{M}$ and the kinetic luminosity $\dot{E_k}$ of the outflow:

    \begin{equation} \label{kinlum}
    \dot{M} = \frac{M}{t_{dyn}} = 4\pi\Omega R N_H \mu m_p v
    \end{equation}
    \begin{equation*}
    \dot{E_k} = \frac{1}{2} \dot{M} v^2 
    \end{equation*}

    We assume $\Omega$ = 0.2 based on the ratio of quasars that show a \ion{C}{IV} BAL \citep{hewett2003frequency}. Using the velocity at the deepest absorption ($v$ $= -$ $720^{+12}_{-12}$ km s$^{-1}$) as the average velocity of the outflow leads to a mass loss rate of $\dot{M}$ = $33^{+13}_{-12}$ $M_{\odot}$ yr$^{-1}$ and kinetic luminosity $\dot{E_k}$ = $5.5^{+2.1}_{-2.0}$ $\times$ $10^{42}$ erg $s^{-1}$ for the outflow, which is $6.8^{+2.7}_{-2.5}$ $\times$ $10^{-3}$ \% of $L_{Bol}$. This is much smaller than the $\gtrsim$ 0.5 \% required for efficient AGN feedback as determined by \cite{hopkins2010quasar}.

    \section{Revisiting the Keck-HIRES Analysis of Hamann et al. (2001)} \label{sec:HamannAnalysis}

    \cite{hamann2001high} observed 3C 191 using the High Resolution Echelle Spectrometer (HIRES) on the Keck I telescope on Mauna Kea, Hawaii. Using multiple observations adding up to 42,000 secs, they covered the wavelength range $\lambda$ $\sim$ 3530 - 8927 \r{A} with spectral resolution R $\approx$ 45,000 and 3$\sigma$ uncertainties $\lesssim$ 7\% for the final fluxes. Using the emission line redshift of z = 1.956 determined by \cite{tytler1992systematic}, they identified an outflow system covering a velocity range $-$1400 $\lesssim$ v $\lesssim$ $-$400 km s$^{-1}$, with three components: $-$1400 $\lesssim$ $\Delta v_1$ $\lesssim$ $-$1160 km s$^{-1}$, $-$1160 $\lesssim$ $\Delta v_2$ $\lesssim$ $-$810 km s$^{-1}$ and $-$810 $\lesssim$ $\Delta v_3$ $\lesssim$ $-$400 km s$^{-1}$. If we instead use the redshift determined by \cite{zhao3c191} (z = 1.9527), these components would have the velocities: $-$1060 $\lesssim$ $\Delta v_1$ $\lesssim$ $-$820 km s$^{-1}$,$-$820 $\lesssim$ $\Delta v_2$ $\lesssim$ $-$470 km s$^{-1}$ and $-$470 $\lesssim$ $\Delta v_3$ $\lesssim$ $-$60 km s$^{-1}$. The outflow identified in the VLT/X-shooter spectrum (with $-$950 $\lesssim$ v $\lesssim$ $-$500 km s$^{-1}$) is thus consistent with their system 2, which is their main outflow component, showing troughs from all species (see their Fig. 2 and Table 1). We can see their system 1 and 3 in the blue and red wings respectively of our troughs, but our Gaussian modeling is able to isolate the main component (see Fig. \ref{fig:Al} and \ref{fig:siII}), thus allowing us to compare the results of their analysis with ours.  
    
    \cite{hamann2001high} analysed the outflow in great detail and estimated its radial distance to be $R \approx 28$ kpc from the central source. They determined log ($U$) $\approx -2.8$ based on the comparison of the ratio of the \ion{Al}{ii}/\ion{Al}{iii} column densities to the theoretical ionization fractions in optically thin photoionized cloud using \textsc{Cloudy}. They also utilized the ratio of the \ion{Si}{ii} $\lambda 1527$ \r{A} and \ion{Si}{ii*} $\lambda 1533$ \r{A} to obtain an estimate of the electron number density $n_e \approx 300$ cm$^{-3}$ and assumed $n_{H} = n_{e}$. Finally, they modeled the ionizing spectrum as a power law with $L_{\nu} \propto \nu^{\alpha}$, where $\alpha = -1.6$ for frequencies above the Lyman limit, and $\alpha = -0.7$ below it (the orange curve in Fig. \ref{fig:SED}). Then they obtained the radial distance of the outflow using their equation (7):
    \begin{equation} \label{distance}
    R = \sqrt{\frac{-L_{LL}}{4\pi c h n_H U \alpha}}
    \end{equation}
    where $\alpha = -1.6$, $h$ is the Planck constant and $L_{LL}$ is the luminosity density at the Lyman Limit in the quasar rest frame, which was estimated by extrapolation of the data to the Lyman limit using the prescribed $\alpha = -0.7$, with $L_{LL} \approx$ 3.6 $\times 10^{31}$ ergs s$^{-1}$ Hz$^{-1}$. This results in:

    \begin{equation} \label{R_calc}
    R = \sqrt{\frac{3.6 \times 10^{31}}{4\pi \cdot 2.998 \times 10^{10} \cdot 6.626 \times 10^{-27} \cdot 300 \cdot 10^{-2.8} \cdot 1.6}}
    \end{equation}
    where all the values are in cgs units. This leads to $R \approx$ 1.38 $\times$ 10$^{23}$ cm = 44.6 kpc. This is 1.6 times larger than their reported value of $R$ $\approx$ 28 kpc obtained using the same parameters. Both these values are off by a factor of several when compared to the distance determined from our analysis of the VLT/X-shooter spectrum ($R \approx$ 5 kpc), and therefore we take a detailed look at this discrepancy in the following sections. 

    \subsection{Change in Continuum and Extrapolation to Lyman Limit} \label{HamannSED}
    We compare the SED used by \cite{hamann2001high} with the scaled HE0238 SED used in our analysis in Fig. \ref{fig:SED}. This reveals that their $\nu L_{\nu}$ is roughly 6 times larger around the Lyman Limit. To understand this difference, we compare the observed spectrums obtained from both Keck/HIRES and VLT/X-shooter. At the longest available observed wavelength from their Fig. 1, $\lambda_{obs} \simeq$ 4976 \r{A}, the measured flux density from the Keck/HIRES spectrum is $F_{\lambda} \simeq$ 2.16 $\times$ 10$^{-16}$ and from the VLT/X-shooter spectrum is 0.91 $\times$ 10$^{-16}$ [ergs cm$^{-2}$ s$^{-1}$ $\r{A}^{-1}$]. A similar factor of increase in flux is also seen between the two spectrum at $\lambda_{obs} \simeq$ 4000 \r{A}. Hence, while there is evidence for a change in continuum between these two epochs, it is much smaller than the difference seen in the SEDs. A likely reason for this could be the interpolation performed by \cite{hamann2001high} from their flux measurement at lower frequencies to the expected flux at Lyman Limit. The Keck/HIRES spectrum is limited on the short wavelength end by $\lambda_{rest} \gtrsim$ 1160 \r{A} and therefore their interpolation to shorter wavelengths is based on their power law index as discussed in the previous section. If we assume that the factor of increase in the continuum remains the same over the wavelength range, we can obtain an estimate for the Keck/HIRES flux at $\lambda_{obs}$ = 10,000 \r{A} as $F_{\lambda} \simeq$ 1.09 $\times$ 10$^{-16}$ ergs cm$^{-2}$ s$^{-1}$ $\r{A}^{-1}$. Using this to scale their SED gives us an estimate for $L_{LL}$ for the Keck/HIRES spectrum, with $L_{LL} \approx$ 1.4 $\times 10^{31}$ ergs s$^{-1}$ Hz$^{-1}$. This is roughly 40\% of the estimate obtained by \cite{hamann2001high}. The SED using the rescaled luminosity density is shown in Fig. \ref{fig:SED} in black.

    \subsection{Photoionization modeling} \label{sec:Hamann_Photoion}

\begin{figure}[h]
   \centering
   \resizebox{\hsize}{!}
        {\includegraphics[width=1.00\linewidth]{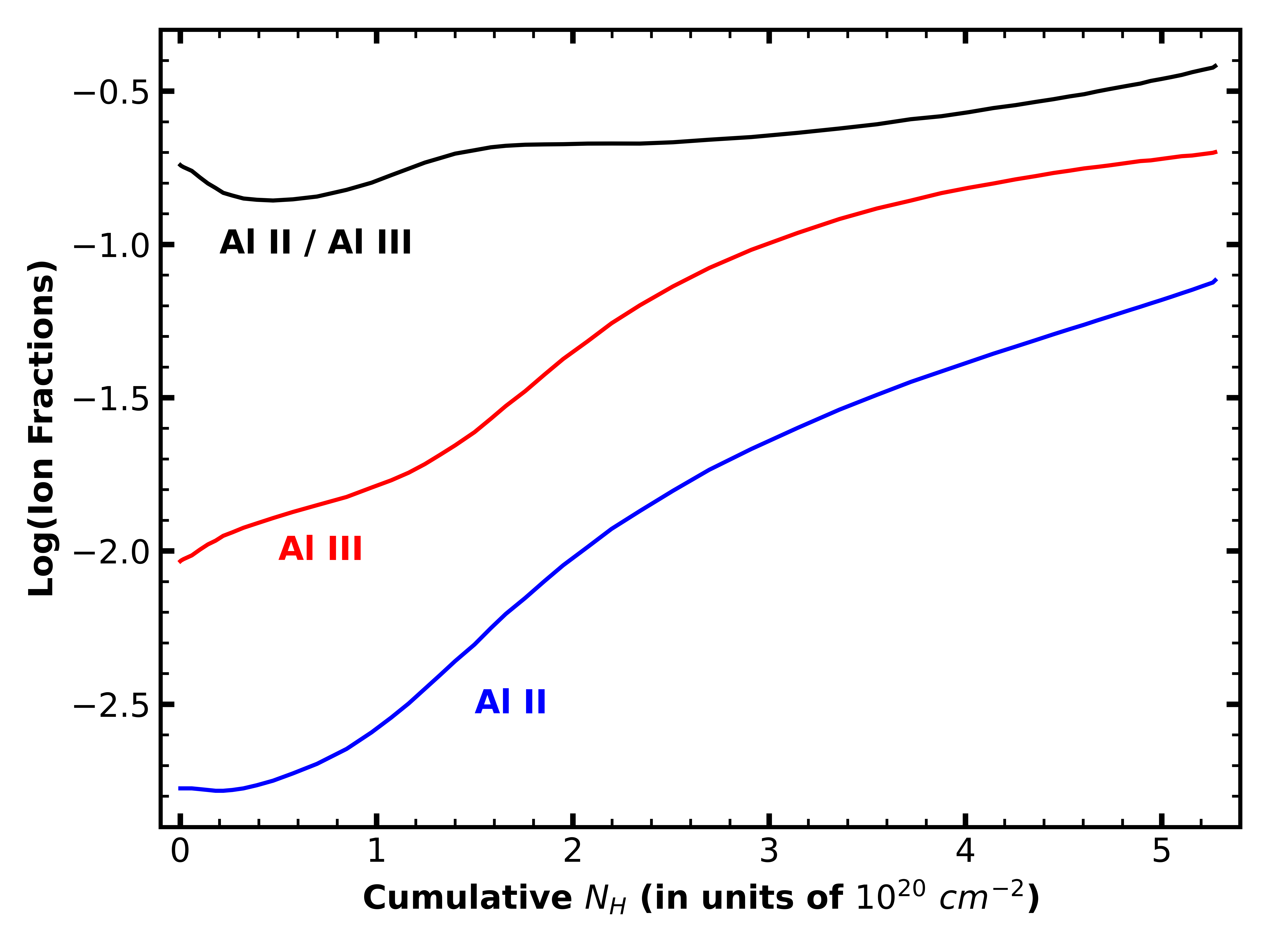}}
      \caption{Ionic fraction within the cloud for log($U_H$) = -2.09. The blue and red curves shows the ionic fraction of \ion{Al}{II} and \ion{Al}{iii} respectively, while the black curve shows their ratio.
         \label{fig:ion_frac}}
   \end{figure}

    Using the methodology described in section \ref{photoion} we can obtain a photoionization model of the outflow. \cite{hamann2001high} determined a covering factor $C_f$ = 0.7 for weak and intermediate lines, which they applied to the Al troughs to obtain their true column densities. They report the corrected total column densities for \ion{Al}{II} and \ion{Al}{III} integrated over all 3 velocity components, but we are interested primarily in their component 2 for which they do not report the corrections explicitly. However, we can note from their Fig. 2 that the \ion{Al}{II} 1671 \r{A} and \ion{Al}{III} 1863 \r{A} troughs have similar residual intensities at their deepest point with $I_r \approx$ 0.25. Therefore any covering factor correction would affect them in the same way and would thus not significantly change their ratios, which is the indicator of the ionization state of the outflow. Therefore, we use their determined column densities of log($N_{ion}$) = 13.4 and 14.0 [cm$^{-2}$] for \ion{Al}{ii} $\lambda$ 1671 \r{A} and \ion{Al}{iii} $\lambda$ 1863 \r{A} respectively for component 2, with an assumed error of $\pm$ 0.1 dex based on the number of significant digits reported. We take the shape of the radiation incident upon the cloud to be the same as the power law SED of \cite{hamann2001high} and obtain the best-fit solution with log $N_H$ = $20.72^{+0.19}_{-0.20}$ [cm$^{-2}$] and log $U_H$ = $-2.09^{+0.18}_{-0.19}$. This $N_H$ is roughly 1.7 times higher than our estimate from the VLT/X-shooter spectrum of log $N_H$ = $20.50^{+0.13}_{-0.16}$ [cm$^{-2}$]. The $U_H$ based on their measurement is about 7 \% lower than our estimate of log $U_H$ = $-2.06^{+0.12}_{-0.14}$ from the VLT spectrum, and is thus the same within the error bars. However, we note that this is $\sim$ 5 times larger than the $U$ determined by \cite{hamann2001high}, who found log($U$) $\approx -$ 2.8. We perform \textsc{Cloudy} simulations for an optically thin cloud and find that the ratio of \ion{Al}{ii} and \ion{Al}{iii} matches the observed ratio of column densities for log($U$) = $-$2.8. However, we also find that the assumption of the cloud being optically thin doesn't hold throughout our model. The ionic fractions for \ion{Al}{II}, \ion{Al}{III} and their ratio for a \textsc{Cloudy} model with log $N_H$ = $20.72$ [cm$^{-2}$] and log $U_H$ = $-2.09$ are shown in Fig. \ref{fig:ion_frac}. We find that the optical depth at the Lyman continuum $\tau_{LC}$ becomes greater than 1 for $N_H \gtrsim$ $10^{20}$ cm$^{-2}$ and the cloud is no longer optically thin. The ratio of the \ion{Al}{ii} and \ion{Al}{III} ionic fractions therefore no longer remains constant, but instead increases as we approach the end of the cloud (the black curve in Fig. \ref{fig:ion_frac}). As most of the contribution to the \ion{Al}{II} and \ion{Al}{III} population comes from these last zones, the average ratio for the cloud is significantly higher than that predicted by the assumption of the entire cloud being optically thin. Therefore the observed ratio of $N_{\ion{Al}{II}} / N_{\ion{Al}{III}} \sim$ $10^{-0.6}$ is reproduced by a cloud with ionization parameter log $U_H$ = $-2.09$. The cloud with ionization parameter log $U_H$ = $-2.80$ on the other hand predicts $N_{\ion{Al}{II}} / N_{\ion{Al}{III}} \sim$ $10^{0.2}$, which is contradicted by the observations. 

    \subsection{Electron Number Density}

\begin{figure}[h]
   \centering
   \resizebox{\hsize}{!}
        {\includegraphics[width=1.00\linewidth]{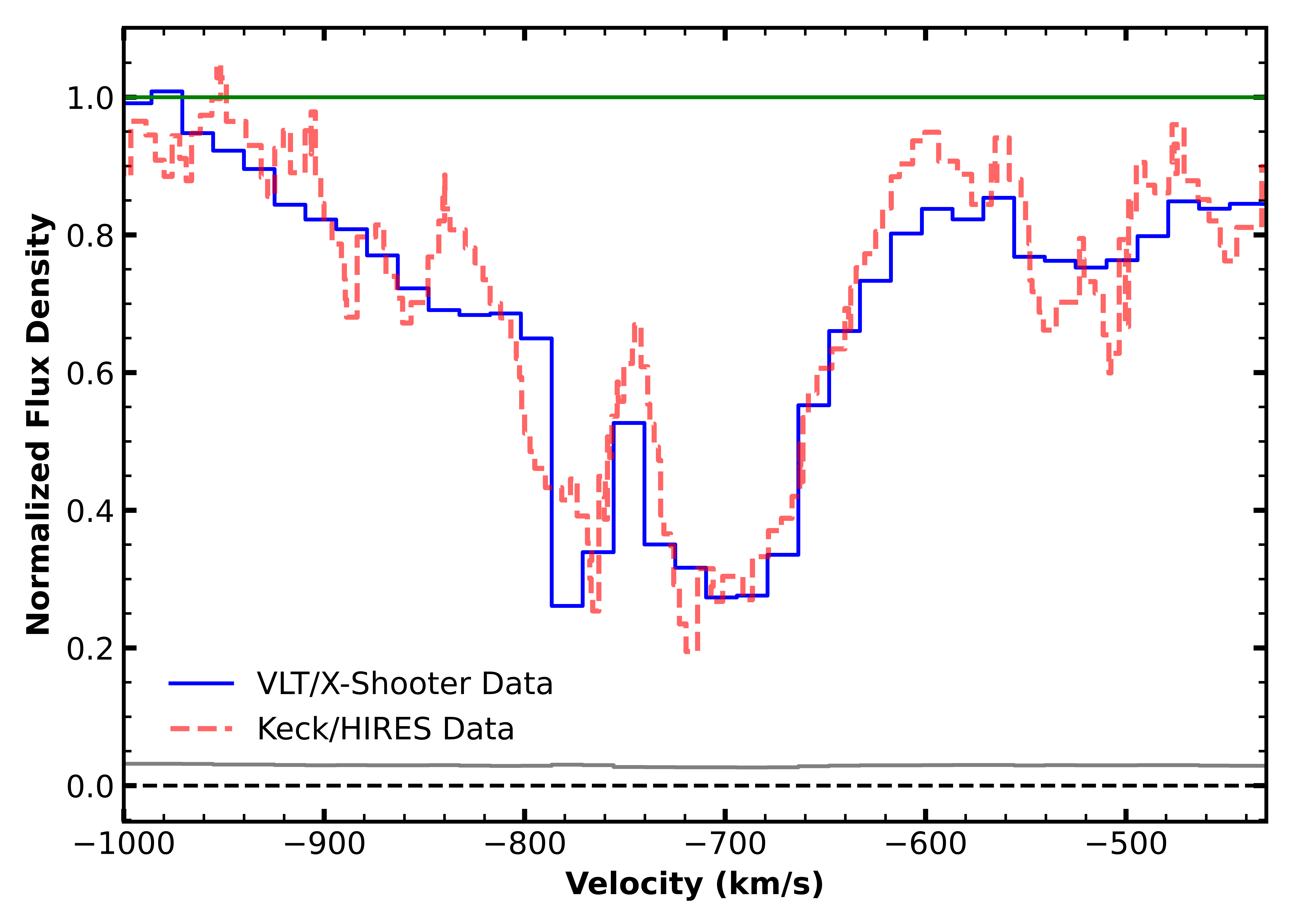}}
      \caption{Comparison between the \ion{Si}{II} 1527 \r{A} trough obtained from the VLT/X-shooter spectrum (blue) and the Keck/HIRES spectrum (red). The Keck/HIRES spectrum was shifted in velocity space to match the troughs for this comparison. The green line shows the local continuum level, and the gray line shows the noise in the VLT/X-shooter flux.}
         \label{fig:troughs}
   \end{figure}

    \cite{hamann2001high} obtained the ratio of the \ion{Si}{ii} 1527 \r{A} and \ion{Si}{ii*} 1533 \r{A} to get an estimate of the electron number density as $n_e \approx 300$ cm$^{-3}$, assuming a covering fraction $C_f$ = 0.7. However we detect a much deeper \ion{Si}{II} 1260 \r{A} trough in the VLT/X-shooter spectrum, which indicates a higher covering fraction for \ion{Si}{ii}. At its deepest point, the \ion{Si}{II} 1260 \r{A} trough has residual intensity $I_v/I_0$ $\sim$ 0.1. Equation (2) from \cite{hamann2001high} would thus suggest $C_f \geq$ 0.9 for \ion{Si}{II}. Such high values of $C_f$ would make a constant covering factor correction negligible for the \ion{Si}{II} troughs and therefore we consider $C_f$ = 1. We also compare the \ion{Si}{ii} 1527 \r{A} and \ion{Si}{ii*} 1533 \r{A} troughs between the two epochs and find little variability for these troughs (see Fig. \ref{fig:troughs}). We note that the \ion{Si}{II} troughs seen in the Keck Spectrum were found to have a higher velocity than the VLT/X-Shooter spectrum by $\sim$ 220 km s$^{-1}$. The change in redshift from $z$ = 1.956 to $z$ = 1.9527 should result in a velocity shift of $\sim$ 340 km s$^{-1}$. Therefore, if we use the same redshift, there is an offset of $\sim$ 120 km s$^{-1}$ between the two spectra, which could be due to absolute wavelength calibration issues with the Keck spectrum.
    
    Due to the lack of any significant variation, we can use our column density measurements for these troughs as they are the same as \cite{hamann2001high}'s determination to the third significant digit, and allow for error estimation based on our modeling, with log($N_{\ion{Si}{II}}$) = $14.59_{-0.02}^{+0.02}$ [cm$^{-2}$] and log($N_{\ion{Si}{II*}}$) = $14.31_{-0.04}^{+0.04}$ [cm$^{-2}$]. Following the steps described in section \ref{sec:eden}, we match the ratio of the column densities to the theoretical prediction from \textsc{Chianti} and obtain $n_e$ = $575^{+85}_{-74}$ cm$^{-3}$. This is in agreement with $n_e \approx$ 510 cm$^{-3}$ determined by \cite{hamann2001high} for $C_f$ = 1. 
    It is also in agreement with our determination of $n_e$ = $602^{+121}_{-124}$ cm$^{3}$ from the \ion{Si}{ii} 1304 \r{A} and \ion{Si}{ii*}  1309 \r{A} troughs in the VLT/X-shooter spectrum.\\
    Finally, We run a \textsc{Cloudy} model for the outflow with the photoionization parameters obtained in section \ref{sec:Hamann_Photoion} and find that the hydrogen ionization front is not entirely developed, similar to our finding from the analysis of the VLT/X-shooter spectrum. Thus we find that averaging over the cloud based on the zone-wise \ion{Si}{ii} column density again yields $n_e\approx$ 1.1 $n_H$.

   \subsection{Distance Determination} \label{dist:Ham}

    We have determined some changes in the values obtained for $L_{LL}$, $U$ and $n_e$ from the Keck/HIRES spectrum in our analysis so far in section \ref{sec:HamannAnalysis}. Using these updated values with eqn. (\ref{distance}), we find that the outflow is located $9.3^{+1.8}_{-2.5}$ kpc away from the central source. This distance is $\sim$ 79 \% smaller than the earlier determination of R $\approx$ 44.6 kpc based on the parameters reported in \cite{hamann2001high}. \par
    However, as we discussed in section \ref{sec:distance}, the HE0238 SED is a better representative of the shape of the radiation incident upon the cloud. Therefore we scale it to the expected Keck/HIRES flux at $\lambda_{obs}$ = 10,000 \r{A} determined in section \ref{HamannSED}. This leads to $Q_H$ = $1.1$ $\times$ $10^{57}$ $s^{-1}$. We also use the HE0238 SED in the photoionization modeling with the Al column densities determined by \cite{hamann2001high} and find that it decreases the required ionization parameter to log $U_H$ = $-2.13^{+0.16}_{-0.18}$. Using these updated values with eqn. (\ref{distance_VLT}) puts the outflow at a distance $R = 8.9^{+1.4}_{-2.7}$ kpc away from the central source. This updated analysis brings the distance determined from the Keck/HIRES spectrum much closer to our estimate of $5.1^{+0.9}_{-1.0}$ kpc based on the VLT/X-shooter spectrum. The difference between these two distance estimates stems mostly from the difference in $Q_H$ which is due to the flux difference between the two spectra. If the Keck/HIRES spectrum is scaled to match the continuum level of the VLT/X-Shooter spectrum, the distance of the outflow determined from the Keck/HIRES spectrum would have been $R = 5.8^{+1.2}_{-1.5}$ kpc, which is in good agreement with our distance determination from VLT/X-shooter spectrum. We discuss the issues concerning this flux difference in section \ref{sec:var}.

    \section{Comparison with the IFS analysis} \label{IFSZhao}

    3C 191 was also observed by the Spectrograph for INtegral Field Observations in the Near Infrared (SINFONI) on VLT in multiple sessions between December 2017 and March 2018 (as part of program: 097.B-0570(B), PI: Benn) with a total integration time of 4800 s. The observations covered a rest-frame spectral range of 4910-6266 \r{A}, with spectral resolution of R = 3000, and angular resolution of $\ang{;;0.24} \times \ang{;;0.24}$, which corresponds to a spatial resolution of $\sim$ 2 $\times$ 2 kpc at the redshift of the quasar. In our companion paper \citep{zhao3c191} we analysed the ionized outflow in emission as traced by the [\ion{O}{iii}] $\lambda \lambda$ 4959, 5007 doublet. The [\ion{O}{iii}] line profile is decomposed into three distinct velocity components: blueshifted, zero velocity and redshifted with respect to the quasar. The zero velocity component is consistent with a typical narrow line region for the quasar, while the blueshifted/redshifted components show strong signs of a bipolar outflow \citep[see section 4 of][for a detailed description of the extended emission line features]{zhao3c191}. For comparison with the outflow detected in absorption, we focus on the blueshifted component of the emission outflow. \par

\begin{figure}[h]
   \centering
   \resizebox{\hsize}{!}
        {\includegraphics[width=0.8\linewidth]{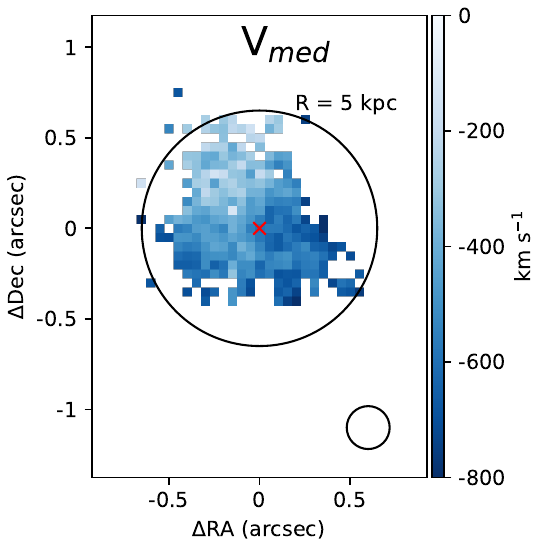}}
      \caption{Median velocity map of the blueshifted component of the ionized outflow seen in emission. The red cross marks the position of the quasar and the smaller open circle depicts the point-spread function (PSF). The $R$ = 5 kpc boundary centered around the quasar is shown by the bigger black circle.}
         \label{fig:emission}
   \end{figure}

    The blueshifted outflow component seen in emission covers a velocity range $-$1090 $\lesssim v \lesssim$ $-$20 km s$^{-1}$. This is consistent with the outflow seen in absorption in the VLT/X-shooter spectrum spanning velocity range $-$950 $\lesssim$ v $\lesssim$ $-$250 km s$^{-1}$. This includes the main absorption system analysed in this paper and its blue wing as seen in Figs. \ref{fig:Al} and \ref{fig:siII}. This kinematic correspondence between the outflows seen in absorption and emission thus suggests a common origin for them. Furthermore, the projected distance determined from the IFS mapping for the blueshifted outflow component is $R \sim$ 5 kpc (Fig. \ref{fig:emission}). This is in remarkable agreement with the indirect distance determination of $R$ = $5.1^{+0.9}_{-1.0}$ kpc obtained for the outflow based on the absorption analysis detailed in section \ref{sec:distance}. This is the first instance where independent emission and absorption analysis of a quasar outflow have yielded consistent values for the distance of the outflow from the central source. 
    While comparing the distances between our absorption and IFS analyses, it is important to note that they do not inherently measure the same distance. The absorption analysis measures the distance of the outflowing component from the central source along our line-of-sight, whereas the IFS analysis measures the projected distance in the sky plane. The relation between these two distance depends upon the geometry of the outflow. Fig. \ref{fig:emission} (along with others in \cite{zhao3c191}) hints towards a roughly spherically symmetric outflow, in which case the line-of-sight distance and the projected distance in the sky plane would be similar. However, other geometries for the outflow (such as bi-polar) cannot be clearly ruled out on the basis of Fig. \ref{fig:emission} alone.

    \cite{zhao3c191} also estimate the energetics for the IFS manifestation of the outflow (see their section 5.2 for a detailed description). The mass of the ionized gas is estimated using the [\ion{O}{III}] luminosity, using the $n_e$ determined from our absorption analysis.  The mass loss rate ($\dot{M}$) and the kinetic luminosity ($\dot{E}_{kin}$) are then calculated accordingly, with the results summarized in Table \ref{table:energetics}. The energetics from the absorption analysis and the IFS analysis are consistent within a factor of 3, with the difference arising mostly due to the smaller $M_{gas}$ obtained for the IFS analysis. We note that the ionized gas mass obtained from the emission lines suffer from greater uncertainties than their absorption counterpart due to several factors, such as the temperature dependence of line emissivities. Moreover, as \cite{carniani2015ionised} show, the ionized gas mass traced by [\ion{O}{III}] is usually lower than that traced by \ion{H}{$\beta$}, which was not covered in the observations analysed by \cite{zhao3c191}. 

   \begingroup
\setlength{\tabcolsep}{12pt}
\renewcommand{\arraystretch}{1.5}
\begin{table}
\caption{Kinematics and Energetics for the outflow in 3C 191}
\centering
\begin{tabular}{lcc}
\hline
& Absorption & IFS \\
\hline
Parameter & Value & Value \\
\hline
$v$ (km s$^{-1}$) & $-$950 to $-$500 & $-$1090 to $-$20   \\
$R$ (kpc) & $5.1^{+0.9}_{-1.0}$ & $5^{+1.3}_{-0.2}$  \\
$t_{dyn}$ (yrs) & $6.9^{+1.2}_{-1.3}$ $\times$ $10^6$ & $5.0^{+0.3}_{-0.3}$ $\times$ $10^6$ \\
$M_{gas}$ ($M_{\odot}$) & $2.3^{+1.1}_{-1.1}$ $\times$ $10^8$ & $6.6^{+0.7}_{-0.7}$ $\times$ $10^7$ \\
$\dot{M}$ ($M_{\odot}$ yr$^{-1}$) & $33^{+13}_{-12}$ & 9.5-13.4  \\
$E_{kin}$ (erg) & $1.2^{+0.6}_{-0.6}$ $\times$ $10^{57}$ & $5.3^{+0.9}_{-0.9}$ $\times$ $10^{56}$ \\
$\dot{E}_{kin}$ (erg s$^{-1}$ ) & $5.5^{+2.1}_{-2.0}$ $\times$ $10^{42}$ & 2.6-3.7 $\times$ $10^{42}$ \\

\hline
\end{tabular}
\tablecomments{The reported values for the absorption analysis are only for the main component and does not include contribution from the associated absorption in the blue wing. For the IFS analysis, the reported kinematics are for the blueshifted component, whereas the energetics correspond to the blueshifted + redshifted component that make up the outflowing bubble.}
\label{table:energetics}
\end{table}
\endgroup

    \section{Discussion} \label{sec:discussion}
    
    \subsection{Variability of the central source} \label{sec:var}

    As discussed in section \ref{dist:Ham}, the difference between the distance estimates obtained from the Keck/HIRES (1997 epoch) and the VLT/X-shooter (2013 epoch) analyses can be traced solely to a factor of $\sim$ 2.4 flux difference between the two spectra. During this period, 3C 191 was also observed by the Catalina Real-time Transient Survey \citep[CRTS;][]{drake2009first} between April 2005 and October 2013. It was also observed later by the Zwicky Transient Facility \citep[ZTF;][]{bellm2018zwicky, malanchev2023snad} between March 2018 and February 2023. Over the combined $\sim$ 15 years of observation, the quasar showed little variability in flux ($\lesssim$ 20\%) in all the observed bands. 
    
    Moreover, equation (\ref{uh}) suggests that variation in the ionizing flux should lead to a change in the ionization state of the gas given that the timescale of flux variation is comparable to the recombination timescale of the gas. For an absorber in photoionization equilibrium, if there is a sudden change of order unity in the incident ionizing flux, the timescale for the change in the ionic fraction of a species is given by \citep{krolik1996observable, arav2012multiwavelength}:

    \begin{equation} \label{timescale}
        t^{*} = \left[ \alpha_i n_e \left(\frac{n_{i+1}}{n_i}-\frac{\alpha_{i-1}}{\alpha_i}\right) \right]^{-1},
    \end{equation}
    where $n_e$ is the electron number density of the gas, $\alpha_i$ is the recombination rate of the ion $i$ and $n_i$ is the fraction of a given element in ionization stage $i$. For the last zone of our modeled cloud, we can obtain $n_e$, $n_i$ and $\alpha_i$ from \textsc{Cloudy}. The timescales for change in ionic fractions of \ion{Si}{II}, \ion{Al}{II} and \ion{Al}{III} can then be determined using eqn. (\ref{timescale}), resulting in $t^*_{\ion{Si}{ii}}$ $\sim$ 1 yr, $t^*_{\ion{Al}{ii}}$ $\sim$ 3 yrs and $t^*_{\ion{Al}{iii}}$ $\sim$ 31 yrs. This suggests that a change in the ionizing flux from the quasar over a timescale $t$ $\gtrsim$ 1 yr would have led to a change in the observed troughs of \ion{Si}{II} and \ion{Al}{II}. However as shown in Fig. \ref{fig:troughs}, no significant change is seen in these troughs between the two epochs. 
    This is consistent with the observations from CRTS and ZTF, and therefore the large difference in the flux seen between the Keck/HIRES and VLT/X-shooter spectra is unlikely to be a result of intrinsic flux variation in the quasar. There could thus be possible issues with the absolute flux-calibration for these observations.

    \subsection{Comparison with other outflows} \label{compareothers}

    \cite{liu2015integral} obtained the IFS maps of two Seyfert 1.5 galaxies: IRAS F04250-5718 and Mrk 509 using the Gemini Multi-Object Spectrograph (GMOS) on the Gemini South telescope (program ID: GS-2013B-Q-84, PI: D. Rupke). Both these objects were already known to have shown outflows in absorption for which indirect distance constraints were available. \cite{liu2015integral} detected ionized outflows in emission for both objects as traced by the [\ion{O}{iii}] $\lambda \lambda$ 4959, 5007 doublet, and thus presented one of the first such comparisons between AGN outflows analysed using absorption and IFS techniques. Here we summarize the results of these analyses in brief.

    IRAS F04250-5718: \cite{liu2015integral} detected a bi-conical outflow extending out to $\sim$ 2.2 kpc on one side and $\sim$ 2.9 kpc on the other. They determined the maximum line-of-sight velocity of the the outflowing gas to be $\sim$ 330 $\pm$ 30 km s$^{-1}$ based on modifications to the spherical outflow model of \cite{liu2013observations}, in which the opening angle of the outflow is measured to be $\sim$ \ang{70;;} as per their IFS map of [\ion{O}{III}] linewidth. Analysis of the outflow seen in absorption in this object was performed by \cite{edmonds2011galactic} based on high-resolution data obtained from the Cosmic Origins Spectrograph (COS) on the Hubble Space Telescope (HST). They identified three different kinematic components spanning velocity range $-$290 $\lesssim$ v $\lesssim$ +30 km s$^{-1}$. Based on the non-detection of excited states from \ion{C}{II}, they derived an upper limit on the electron number density of the outflow with $n_e \lesssim$ 30 cm$^{-3}$. This provided a constraint on the distance of the outflow from the central source and placed it at-least 3 kpc away ($R \gtrsim$ 3 kpc). This is consistent with the spatial extent of the outflow on one side of the quasar.

    Mrk 509: The outflow detected from the IFS mapping by \cite{liu2015integral} shows a roughly spherical morphology with a radius of $\sim$ 1.2 kpc, and velocity v $\sim$ 293 $\pm$ 51 km s$^{-1}$. On the absorption front, \cite{kraemer2003kinematics} obtained Space Telescope Imaging Spectrograph (STIS) echelle spectrum of the AGN. Mrk 509 was also the subject of a large multiwavelength campaign \citep{kaastra2011multiwavelength} as part of which it was observed with HST/COS. \cite{kriss2011multiwavelength} identified nine absorption systems spanning velocity range  $-$425 $\lesssim$ v $\lesssim$ +250 km s$^{-1}$. Comparison between the \ion{C}{IV} and \ion{N}{V} troughs seen in the STIS and COS spectrums showed negligible variation despite a large change in the ionizing flux. This allowed them to obtain upper limits on the electron number densities based on recombination timescale arguments. Using this, they determined a lower limit on the distance to the absorber of 100-200 pc from the ionizing source for all the components, which is again consistent with the spatial extent of the outflow determined from the IFS mapping ($\sim$ 1.2 kpc). \par

    Therefore, the IFS analysis of both these objects revealed that the spatial extent of the outflows were consistent with the constraints determined from the indirect determination. However, as the absorption analysis only provided lower limits on the indirect distance determination, a conclusive correspondence can not be drawn between the distances obtained from the two techniques. Therefore, a direct quantitative comparison of the absorption and emission techniques is featured for the first time in this work and our companion paper \cite{zhao3c191}.
    
    \section{Summary} \label{summary}

    This paper (along with the companion paper by \citealt{zhao3c191}) presents a detailed study of the ionized outflow associated with the radio-loud quasar 3C 191 using absorption and IFS analysis techniques. In the VLT/X-Shooter spectrum of the quasar, we identify an outflow system in absorption with velocity v $\sim -$ 720 km s$^{-1}$. We detect multiple ionized species covering a broad range of ionization and model the observed troughs to obtain their column densities (Section \ref{sec:colden}). We then obtain the best fit photoionization model for the cloud with total hydrogen column density log $N_H$ = $20.50_{-0.16}^{+0.13}$ [cm$^{-2}$] and ionization parameter log $U_H$ = $-2.06_{-0.14}^{+0.12}$ based on the observed \ion{Al}{II} and \ion{Al}{iii} column densities (Section \ref{photoion}). The detection of the metastable excited state of \ion{Si}{II} allows us to obtain the electron number density of the outflow with log $n_e$ = $2.78_{-0.10}^{+0.08}$ [cm$^{-3}$] (Section \ref{sec:eden}). Combining these parameters with our informed assumption for the incident SED (Fig. \ref{fig:SED}) locates the outflow at a distance of $R$ = $5.1^{+0.9}_{-1.0}$ kpc from the central source. This leads to a mass loss rate of $\dot{M}$ = $33^{+13}_{-12}$ $M_{\odot}$ yr$^{-1}$ and kinetic luminosity $\dot{E_k}$ = $5.5^{+2.1}_{-2.0}$ $\times$ $10^{42}$ erg $s^{-1}$ for the outflow. \par
    We note that our distance determination from the VLT/X-Shooter spectrum is in disagreement with that of \cite{hamann2001high} who analysed the Keck/HIRES spectrum of 3C 191 and placed the outflow at a much larger distance of $R \sim$ 28 kpc. We revisit their analysis in section \ref{sec:HamannAnalysis} and obtain different values for the number of ionizing photons ($Q_H$) and the ionization parameter ($U_H$) based on our analysis techniques detailed in section \ref{HamannSED} and \ref{sec:Hamann_Photoion} respectively. This leads to an updated value for the distance determined from the Keck/HIRES spectra with $R = 8.9^{+1.4}_{-2.7}$ kpc, which is much closer to our estimate from the VLT/X-Shooter spectrum. 
    \par
    \cite{zhao3c191} details our analysis of the spatially-resolved integral field observations of the ionized outflow surrounding 3C 191 obtained using VLT/SINFONI. In section \ref{IFSZhao}, we compare the properties of the [\ion{O}{III}] outflow seen in emission with the outflow detected in absorption. These two independent approaches lead to a remarkably consistent picture for the outflow as the distance determined indirectly using absorption analysis is verified by the spatial extent of the outflow determined directly from the IFS observations for the first time for a quasar outflow. The kinematics and energetics of the two manifestations of the outflow are also found to be in agreement as shown in Table \ref{table:energetics}. 
    \begin{acknowledgements}
    We acknowledge support from NSF grant AST 2106249, as well as NASA STScI grants AR-15786, AR-16600, AR-16601 and AR-17556. Q.Z., L.S. and G.L. acknowledge the support from the China Manned Space Project (the second-stage CSST science project: {\em Investigation of small-scale structures in galaxies and forecasting of observations}, Nos. CMS-CSST-2021-A06 and CMS-CSST-2021-A07), the National Natural Science Foundation of China (No. 12273036), the Ministry of Science and Technology of China (No. 2023YFA1608100), and the Cyrus Chun Ying Tang Foundations. Q.Z. and J.W. acknowledge National Natural Science Foundation of China (NSFC) grants 12033004, 12333002. Q.Z. acknowledges the support from the China Postdoctoral Science Foundation (2023M732955). We also thank the anonymous referee for their constructive comments that helped improve this paper.
    \end{acknowledgements}

\bibliography{3C191}{}

\newcommand{\noop}[1]{}
\begin{thebibliography}{}
\expandafter\ifx\csname natexlab\endcsname\relax\def\natexlab#1{#1}\fi
\providecommand{\url}[1]{\href{#1}{#1}}
\providecommand{\dodoi}[1]{doi:~\href{http://doi.org/#1}{\nolinkurl{#1}}}
\providecommand{\doeprint}[1]{\href{http://ascl.net/#1}{\nolinkurl{http://ascl.net/#1}}}
\providecommand{\doarXiv}[1]{\href{https://arxiv.org/abs/#1}{\nolinkurl{https://arxiv.org/abs/#1}}}

\bibitem[{Akujor {et~al.}(1994)Akujor, L{\"u}dke, Browne, Leahy, Garrington, Jackson, \& Thomasson}]{akujor1994images}
Akujor, C.~E., L{\"u}dke, E., Browne, I., {et~al.} 1994, Astronomy and Astrophysics Suppl., Vol. 105, p. 247-258 (1994), 105, 247

\bibitem[{Alexander {et~al.}(2010)Alexander, Swinbank, Smail, McDermid, \& Nesvadba}]{alexander2010searching}
Alexander, D., Swinbank, A., Smail, I., McDermid, R., \& Nesvadba, N. 2010, Monthly Notices of the Royal Astronomical Society, 402, 2211

\bibitem[{Aoki {et~al.}(2011)Aoki, Oyabu, Dunn, Arav, Edmonds, Korista, Matsuhara, \& Toba}]{aoki2011outflow}
Aoki, K., Oyabu, S., Dunn, J.~P., {et~al.} 2011, Publications of the Astronomical Society of Japan, 63, S457

\bibitem[{Arav {et~al.}(2013)Arav, Borguet, Chamberlain, Edmonds, \& Danforth}]{arav2013quasar}
Arav, N., Borguet, B., Chamberlain, C., Edmonds, D., \& Danforth, C. 2013, Monthly Notices of the Royal Astronomical Society, 436, 3286

\bibitem[{Arav {et~al.}(2005)Arav, Kaastra, Kriss, Korista, Gabel, \& Proga}]{arav2005x}
Arav, N., Kaastra, J., Kriss, G.~A., {et~al.} 2005, The Astrophysical Journal, 620, 665

\bibitem[{Arav {et~al.}(1999)Arav, Korista, de~Kool, Junkkarinen, \& Begelman}]{arav1999hubble}
Arav, N., Korista, K.~T., de~Kool, M., Junkkarinen, V.~T., \& Begelman, M.~C. 1999, The Astrophysical Journal, 516, 27

\bibitem[{Arav {et~al.}(2018)Arav, Liu, Xu, Stidham, Benn, \& Chamberlain}]{arav2018evidence}
Arav, N., Liu, G., Xu, X., {et~al.} 2018, The Astrophysical Journal, 857, 60

\bibitem[{Arav {et~al.}(2020)Arav, Xu, Miller, Kriss, \& Plesha}]{arav2020hst}
Arav, N., Xu, X., Miller, T., Kriss, G.~A., \& Plesha, R. 2020, The Astrophysical Journal Supplement Series, 247, 37

\bibitem[{Arav {et~al.}(2012)Arav, Edmonds, Borguet, Kriss, Kaastra, Behar, Bianchi, Cappi, Costantini, Detmers, {et~al.}}]{arav2012multiwavelength}
Arav, N., Edmonds, D., Borguet, B., {et~al.} 2012, Astronomy \& Astrophysics, 544, A33

\bibitem[{Bahcall {et~al.}(1967)Bahcall, Sargent, \& Schmidt}]{bahcall1967analysis}
Bahcall, J.~N., Sargent, W.~L., \& Schmidt, M. 1967, Astrophysical Journal, vol. 149, p. L11, 149, L11

\bibitem[{Baron \& Netzer(2019)}]{baron2019discovering}
Baron, D., \& Netzer, H. 2019, Monthly Notices of the Royal Astronomical Society, 486, 4290

\bibitem[{Bautista {et~al.}(2010)Bautista, Dunn, Arav, Korista, Moe, \& Benn}]{bautista2010distance}
Bautista, M.~A., Dunn, J.~P., Arav, N., {et~al.} 2010, The Astrophysical Journal, 713, 25

\bibitem[{Bellm {et~al.}(2018)Bellm, Kulkarni, Graham, Dekany, Smith, Riddle, Masci, Helou, Prince, Adams, {et~al.}}]{bellm2018zwicky}
Bellm, E.~C., Kulkarni, S.~R., Graham, M.~J., {et~al.} 2018, Publications of the Astronomical Society of the Pacific, 131, 018002

\bibitem[{Berk {et~al.}(2001)Berk, Richards, Bauer, Strauss, Schneider, Heckman, York, Hall, Fan, Knapp, {et~al.}}]{berk2001composite}
Berk, D. E.~V., Richards, G.~T., Bauer, A., {et~al.} 2001, The Astronomical Journal, 122, 549

\bibitem[{Bischetti {et~al.}(2024)Bischetti, Choi, Fiore, Feruglio, Carniani, D'Odorico, Ba{\~n}ados, Chen, Decarli, Gallerani, {et~al.}}]{bischetti2024multi}
Bischetti, M., Choi, H., Fiore, F., {et~al.} 2024, arXiv preprint arXiv:2404.12443

\bibitem[{Borguet {et~al.}(2012{\natexlab{a}})Borguet, Arav, Edmonds, Chamberlain, \& Benn}]{borguet2012major}
Borguet, B.~C., Arav, N., Edmonds, D., Chamberlain, C., \& Benn, C. 2012{\natexlab{a}}, The Astrophysical Journal, 762, 49

\bibitem[{Borguet {et~al.}(2012{\natexlab{b}})Borguet, Edmonds, Arav, Dunn, \& Kriss}]{borguet201210}
Borguet, B.~C., Edmonds, D., Arav, N., Dunn, J., \& Kriss, G.~A. 2012{\natexlab{b}}, The Astrophysical Journal, 751, 107

\bibitem[{Burbidge {et~al.}(1966)Burbidge, Lynds, \& Burbidge}]{burbidge1966measurement}
Burbidge, E., Lynds, C., \& Burbidge, G. 1966, Astrophysical Journal, vol. 144, p. 447, 144, 447

\bibitem[{Byun {et~al.}(2024{\natexlab{a}})Byun, Arav, Dehghanian, Walker, \& Kriss}]{byun2024bal}
Byun, D., Arav, N., Dehghanian, M., Walker, G., \& Kriss, G.~A. 2024{\natexlab{a}}, Monthly Notices of the Royal Astronomical Society, 529, 3550

\bibitem[{Byun {et~al.}(2022{\natexlab{a}})Byun, Arav, \& Hall}]{byun2022vlt}
Byun, D., Arav, N., \& Hall, P.~B. 2022{\natexlab{a}}, Monthly Notices of the Royal Astronomical Society, 517, 1048

\bibitem[{Byun {et~al.}(2024{\natexlab{b}})Byun, Arav, Sharma, Dehghanian, \& Walker}]{byun2024extreme}
Byun, D., Arav, N., Sharma, M., Dehghanian, M., \& Walker, G. 2024{\natexlab{b}}, Astronomy \& Astrophysics, 684, A158

\bibitem[{Byun {et~al.}(2022{\natexlab{b}})Byun, Arav, \& Walker}]{byun2022vltb}
Byun, D., Arav, N., \& Walker, A. 2022{\natexlab{b}}, Monthly Notices of the Royal Astronomical Society, 516, 100

\bibitem[{Carniani {et~al.}(2015)Carniani, Marconi, Maiolino, Balmaverde, Brusa, Cano-D{\'\i}az, Cicone, Comastri, Cresci, Fiore, {et~al.}}]{carniani2015ionised}
Carniani, S., Marconi, A., Maiolino, R., {et~al.} 2015, Astronomy \& Astrophysics, 580, A102

\bibitem[{Cattaneo {et~al.}(2009)Cattaneo, Faber, Binney, Dekel, Kormendy, Mushotzky, Babul, Best, Br{\"u}ggen, Fabian, {et~al.}}]{cattaneo2009role}
Cattaneo, A., Faber, S., Binney, J., {et~al.} 2009, Nature, 460, 213

\bibitem[{Chamberlain \& Arav(2015)}]{chamberlain2015large}
Chamberlain, C., \& Arav, N. 2015, Monthly Notices of the Royal Astronomical Society, 454, 675

\bibitem[{Chamberlain {et~al.}(2015)Chamberlain, Arav, \& Benn}]{chamberlain2015strong}
Chamberlain, C., Arav, N., \& Benn, C. 2015, Monthly Notices of the Royal Astronomical Society, 450, 1085

\bibitem[{{Chatzikos} {et~al.}(2023){Chatzikos}, {Bianchi}, {Camilloni}, {Chakraborty}, {Gunasekera}, {Guzm{\'a}n}, {Milby}, {Sarkar}, {Shaw}, {van Hoof}, \& {Ferland}}]{cloudy2023}
{Chatzikos}, M., {Bianchi}, S., {Camilloni}, F., {et~al.} 2023, \rmxaa, 59, 327, \dodoi{10.22201/ia.01851101p.2023.59.02.12}

\bibitem[{Czerny {et~al.}(2004)Czerny, Li, Loska, \& Szczerba}]{czerny2004extinction}
Czerny, B., Li, J., Loska, Z., \& Szczerba, R. 2004, Monthly Notices of the Royal Astronomical Society, 348, L54

\bibitem[{Dav{\'e} {et~al.}(2019)Dav{\'e}, Angl{\'e}s-Alc{\'a}zar, Narayanan, Li, Rafieferantsoa, \& Appleby}]{dave2019simba}
Dav{\'e}, R., Angl{\'e}s-Alc{\'a}zar, D., Narayanan, D., {et~al.} 2019, Monthly Notices of the Royal Astronomical Society, 486, 2827

\bibitem[{Davidson \& Netzer(1979)}]{davidson1979emission}
Davidson, K., \& Netzer, H. 1979, Reviews of Modern Physics, 51, 715

\bibitem[{Davies {et~al.}(2020)Davies, Baron, Shimizu, Netzer, Burtscher, De~Zeeuw, Genzel, Hicks, Koss, Lin, {et~al.}}]{davies2020ionized}
Davies, R., Baron, D., Shimizu, T., {et~al.} 2020, Monthly Notices of the Royal Astronomical Society, 498, 4150

\bibitem[{de~Kool {et~al.}(2001)de~Kool, Arav, Becker, Gregg, White, Laurent-Muehleisen, Price, \& Korista}]{de2001keck}
de~Kool, M., Arav, N., Becker, R.~H., {et~al.} 2001, The Astrophysical Journal, 548, 609

\bibitem[{Dehghanian {et~al.}(2024)Dehghanian, Arav, Byun, Walker, \& Sharma}]{dehghanian2024narrow}
Dehghanian, M., Arav, N., Byun, D., Walker, G., \& Sharma, M. 2024, Monthly Notices of the Royal Astronomical Society, 527, 7825

\bibitem[{Dere {et~al.}(1997)Dere, Landi, Mason, Fossi, \& Young}]{dere1997chianti}
Dere, K., Landi, E., Mason, H., Fossi, B.~M., \& Young, P. 1997, Astronomy and Astrophysics Supplement Series, 125, 149

\bibitem[{Dere {et~al.}(2019)Dere, Del~Zanna, Young, Landi, \& Sutherland}]{dere2019chianti}
Dere, K.~P., Del~Zanna, G., Young, P.~R., Landi, E., \& Sutherland, R.~S. 2019, The Astrophysical Journal Supplement Series, 241, 22

\bibitem[{Donnari {et~al.}(2021)Donnari, Pillepich, Joshi, Nelson, Genel, Marinacci, Rodriguez-Gomez, Pakmor, Torrey, Vogelsberger, {et~al.}}]{donnari2021quenched}
Donnari, M., Pillepich, A., Joshi, G.~D., {et~al.} 2021, Monthly Notices of the Royal Astronomical Society, 500, 4004

\bibitem[{Drake {et~al.}(2009)Drake, Djorgovski, Mahabal, Beshore, Larson, Graham, Williams, Christensen, Catelan, Boattini, {et~al.}}]{drake2009first}
Drake, A., Djorgovski, S., Mahabal, A., {et~al.} 2009, The Astrophysical Journal, 696, 870

\bibitem[{Dubois {et~al.}(2012)Dubois, Devriendt, Slyz, \& Teyssier}]{dubois2012self}
Dubois, Y., Devriendt, J., Slyz, A., \& Teyssier, R. 2012, Monthly Notices of the Royal Astronomical Society, 420, 2662

\bibitem[{Dunn {et~al.}(2010)Dunn, Bautista, Arav, Moe, Korista, Costantini, Benn, Ellison, \& Edmonds}]{dunn2010quasar}
Dunn, J.~P., Bautista, M., Arav, N., {et~al.} 2010, The Astrophysical Journal, 709, 611

\bibitem[{Edmonds {et~al.}(2011)Edmonds, Borguet, Arav, Dunn, Penton, Kriss, Korista, Costantini, Steenbrugge, Gonzalez-Serrano, {et~al.}}]{edmonds2011galactic}
Edmonds, D., Borguet, B., Arav, N., {et~al.} 2011, The Astrophysical Journal, 739, 7

\bibitem[{Evans {et~al.}(2010)Evans, Primini, Glotfelty, Anderson, Bonaventura, Chen, Davis, Doe, Evans, Fabbiano, {et~al.}}]{evans2010chandra}
Evans, I.~N., Primini, F.~A., Glotfelty, K.~J., {et~al.} 2010, The Astrophysical Journal Supplement Series, 189, 37

\bibitem[{{Evans} {et~al.}(2020){Evans}, {Primini}, {Miller}, {Evans}, {Allen}, {Anderson}, {Becker}, {Budynkiewicz}, {Burke}, {Chen}, {Civano}, {D'Abrusco}, {Doe}, {Fabbiano}, {Martinez Galarza}, {Gibbs}, {Glotfelty}, {Graessle}, {Grier}, {Hain}, {Hall}, {Harbo}, {Houck}, {Lauer}, {Laurino}, {Lee}, {McCollough}, {McDowell}, {McLaughlin}, {Morgan}, {Mossman}, {Nguyen}, {Nichols}, {Nowak}, {Paxson}, {Perdikeas}, {Plummer}, {Rots}, {Siemiginowska}, {Sundheim}, {Thong}, {Tibbetts}, {Van Stone}, {Winkelman}, \& {Zografou}}]{2020AAS...23515405E}
{Evans}, I.~N., {Primini}, F.~A., {Miller}, J.~B., {et~al.} 2020, in American Astronomical Society Meeting Abstracts, Vol. 235, American Astronomical Society Meeting Abstracts \#235, 154.05

\bibitem[{Finn {et~al.}(2014)Finn, Morris, Crighton, Hamann, Done, Theuns, Fumagalli, Tejos, \& Worseck}]{finn2014compact}
Finn, C.~W., Morris, S.~L., Crighton, N.~H., {et~al.} 2014, Monthly Notices of the Royal Astronomical Society, 440, 3317

\bibitem[{Gaskell {et~al.}(2004)Gaskell, Goosmann, Antonucci, \& Whysong}]{gaskell2004nuclear}
Gaskell, C.~M., Goosmann, R.~W., Antonucci, R.~R., \& Whysong, D.~H. 2004, The Astrophysical Journal, 616, 147

\bibitem[{Hall {et~al.}(2002)Hall, Anderson, Strauss, York, Richards, Fan, Knapp, Schneider, Berk, Geballe, {et~al.}}]{hall2002unusual}
Hall, P.~B., Anderson, S.~F., Strauss, M.~A., {et~al.} 2002, The Astrophysical Journal Supplement Series, 141, 267

\bibitem[{{Hamann} \& {Sabra}(2004)}]{hamann2004asp}
{Hamann}, F., \& {Sabra}, B. 2004, in Astronomical Society of the Pacific Conference Series, Vol. 311, AGN Physics with the Sloan Digital Sky Survey, 203, \dodoi{10.48550/arXiv.astro-ph/0310668}

\bibitem[{Hamann {et~al.}(2001)Hamann, Barlow, Chaffee, Foltz, \& Weymann}]{hamann2001high}
Hamann, F.~W., Barlow, T., Chaffee, F., Foltz, C., \& Weymann, R. 2001, The Astrophysical Journal, 550, 142

\bibitem[{Harrison {et~al.}(2014)Harrison, Alexander, Mullaney, \& Swinbank}]{harrison2014kiloparsec}
Harrison, C., Alexander, D., Mullaney, J., \& Swinbank, A. 2014, Monthly Notices of the Royal Astronomical Society, 441, 3306

\bibitem[{Harrison {et~al.}(2012)Harrison, Alexander, Swinbank, Smail, Alaghband-Zadeh, Bauer, Chapman, Del~Moro, Hickox, Ivison, {et~al.}}]{harrison2012energetic}
Harrison, C., Alexander, D., Swinbank, A., {et~al.} 2012, Monthly Notices of the Royal Astronomical Society, 426, 1073

\bibitem[{He {et~al.}(2022)He, Liu, Wang, Mou, Green, Bian, Wang, Ho, Sun, Shen, {et~al.}}]{he2022evidence}
He, Z., Liu, G., Wang, T., {et~al.} 2022, Science Advances, 8, eabk3291

\bibitem[{Hewett \& Foltz(2003)}]{hewett2003frequency}
Hewett, P.~C., \& Foltz, C.~B. 2003, The Astronomical Journal, 125, 1784

\bibitem[{Hlavacek-Larrondo {et~al.}(2024)Hlavacek-Larrondo, Li, \& Churazov}]{hlavacek2024agn}
Hlavacek-Larrondo, J., Li, Y., \& Churazov, E. 2024, in Handbook of X-ray and Gamma-ray Astrophysics (Springer), 4895--4960

\bibitem[{Holt {et~al.}(2011)Holt, Tadhunter, Morganti, \& Emonts}]{holt2011impact}
Holt, J., Tadhunter, C., Morganti, R., \& Emonts, B. 2011, Monthly Notices of the Royal Astronomical Society, 410, 1527

\bibitem[{Hopkins \& Elvis(2010)}]{hopkins2010quasar}
Hopkins, P.~F., \& Elvis, M. 2010, Monthly Notices of the Royal Astronomical Society, 401, 7

\bibitem[{Kaastra {et~al.}(2011)Kaastra, Petrucci, Cappi, Arav, Behar, Bianchi, Bloom, Blustin, Branduardi-Raymont, Costantini, {et~al.}}]{kaastra2011multiwavelength}
Kaastra, J., Petrucci, P.-O., Cappi, M., {et~al.} 2011, Astronomy \& Astrophysics, 534, A36

\bibitem[{Kakkad {et~al.}(2020)Kakkad, Mainieri, Vietri, Carniani, Harrison, Perna, Scholtz, Circosta, Cresci, Husemann, {et~al.}}]{kakkad2020super}
Kakkad, D., Mainieri, V., Vietri, G., {et~al.} 2020, Astronomy \& Astrophysics, 642, A147

\bibitem[{Karouzos {et~al.}(2016)Karouzos, Woo, \& Bae}]{karouzos2016unraveling}
Karouzos, M., Woo, J.-H., \& Bae, H.-J. 2016, The Astrophysical Journal, 819, 148

\bibitem[{Kim {et~al.}(2023)Kim, Woo, Luo, Chung, Baek, Le, \& Son}]{kim2023unraveling}
Kim, C., Woo, J.-H., Luo, R., {et~al.} 2023, The Astrophysical Journal, 958, 145

\bibitem[{Kraemer {et~al.}(2003)Kraemer, Crenshaw, Yaqoob, McKernan, Gabel, George, Turner, \& Dunn}]{kraemer2003kinematics}
Kraemer, S., Crenshaw, D., Yaqoob, T., {et~al.} 2003, The Astrophysical Journal, 582, 125

\bibitem[{Kriss {et~al.}(2011)Kriss, Arav, Kaastra, Ebrero, Pinto, Borguet, Edmonds, Costantini, Steenbrugge, Detmers, {et~al.}}]{kriss2011multiwavelength}
Kriss, G., Arav, N., Kaastra, J., {et~al.} 2011, Astronomy \& Astrophysics, 534, A41

\bibitem[{Krolik \& Kriss(1996)}]{krolik1996observable}
Krolik, J., \& Kriss, G. 1996, Astrophysical Journal v. 456, p. 909, 456, 909

\bibitem[{Krolik(1999)}]{krolik1999active}
Krolik, J.~H. 1999, Active galactic nuclei: from the central black hole to the galactic environment, Vol.~59 (Princeton University Press)

\bibitem[{Leighly {et~al.}(2024)Leighly, Choi, Eracleous, Terndrup, Gallagher, \& Richards}]{leighly2024physical}
Leighly, K.~M., Choi, H., Eracleous, M., {et~al.} 2024, arXiv preprint arXiv:2402.07855

\bibitem[{Liu {et~al.}(2015)Liu, Arav, \& Rupke}]{liu2015integral}
Liu, G., Arav, N., \& Rupke, D.~S. 2015, The Astrophysical Journal Supplement Series, 221, 9

\bibitem[{Liu {et~al.}(2014)Liu, Zakamska, \& Greene}]{liu2014similarity}
Liu, G., Zakamska, N.~L., \& Greene, J.~E. 2014, Monthly Notices of the Royal Astronomical Society, 442, 1303

\bibitem[{Liu {et~al.}(2013)Liu, Zakamska, Greene, Nesvadba, \& Liu}]{liu2013observations}
Liu, G., Zakamska, N.~L., Greene, J.~E., Nesvadba, N.~P., \& Liu, X. 2013, Monthly Notices of the Royal Astronomical Society, 436, 2576

\bibitem[{Lucy {et~al.}(2014)Lucy, Leighly, Terndrup, Dietrich, \& Gallagher}]{lucy2014tracing}
Lucy, A.~B., Leighly, K.~M., Terndrup, D.~M., Dietrich, M., \& Gallagher, S.~C. 2014, The Astrophysical Journal, 783, 58

\bibitem[{Malanchev {et~al.}(2023)Malanchev, Kornilov, Pruzhinskaya, Ishida, Aleo, Korolev, Lavrukhina, Russeil, Sreejith, Volnova, {et~al.}}]{malanchev2023snad}
Malanchev, K., Kornilov, M.~V., Pruzhinskaya, M.~V., {et~al.} 2023, Publications of the Astronomical Society of the Pacific, 135, 024503

\bibitem[{Miller {et~al.}(2020)Miller, Arav, Xu, Kriss, \& Plesha}]{miller2020hst}
Miller, T.~R., Arav, N., Xu, X., Kriss, G.~A., \& Plesha, R.~J. 2020, The Astrophysical Journal Supplement Series, 247, 41

\bibitem[{Miller {et~al.}(2018)Miller, Arav, Xu, Kriss, Plesha, Benn, \& Liu}]{miller2018distance}
Miller, T.~R., Arav, N., Xu, X., {et~al.} 2018, The Astrophysical Journal, 865, 90

\bibitem[{Modigliani {et~al.}(2010)Modigliani, Goldoni, Royer, Haigron, Guglielmi, Fran{\c{c}}ois, Horrobin, Bristow, Vernet, Moehler, {et~al.}}]{modigliani2010x}
Modigliani, A., Goldoni, P., Royer, F., {et~al.} 2010, in Observatory operations: Strategies, processes, and systems iii, Vol. 7737, SPIE, 572--583

\bibitem[{Moe {et~al.}(2009)Moe, Arav, Bautista, \& Korista}]{moe2009quasar}
Moe, M., Arav, N., Bautista, M.~A., \& Korista, K.~T. 2009, The Astrophysical Journal, 706, 525

\bibitem[{Nesvadba {et~al.}(2008)Nesvadba, Lehnert, De~Breuck, Gilbert, \& Van~Breugel}]{nesvadba2008evidence}
Nesvadba, N., Lehnert, M.~D., De~Breuck, C., Gilbert, A., \& Van~Breugel, W. 2008, Astronomy \& Astrophysics, 491, 407

\bibitem[{Noterdaeme {et~al.}(2021)Noterdaeme, Balashev, Krogager, Laursen, Srianand, Gupta, Petitjean, \& Fynbo}]{noterdaeme2021down}
Noterdaeme, P., Balashev, S., Krogager, J.-K., {et~al.} 2021, Astronomy \& Astrophysics, 646, A108

\bibitem[{Osterbrock \& Ferland(2006)}]{osterbrock2006astrophysics}
Osterbrock, D.~E., \& Ferland, G.~J. 2006, Astrophysics Of Gas Nebulae and Active Galactic Nuclei (University science books)

\bibitem[{{Parlanti} {et~al.}(2024){Parlanti}, {Carniani}, {{\"U}bler}, {Venturi}, {Circosta}, {D'Eugenio}, {Arribas}, {Bunker}, {Charlot}, {L{\"u}tzgendorf}, {Maiolino}, {Perna}, {Rodr{\'\i}guez Del Pino}, {Willott}, {B{\"o}ker}, {Cameron}, {Chevallard}, {Cresci}, {Jones}, {Kumari}, {Lamperti}, \& {Scholtz}}]{parlanti2024}
{Parlanti}, E., {Carniani}, S., {{\"U}bler}, H., {et~al.} 2024, \aap, 684, A24, \dodoi{10.1051/0004-6361/202347914}

\bibitem[{Pearson {et~al.}(1985)Pearson, Perley, \& Readhead}]{pearson1985compact}
Pearson, T., Perley, R., \& Readhead, A. 1985, Astronomical Journal (ISSN 0004-6256), vol. 90, May 1985, p. 738-755., 90, 738

\bibitem[{{Planck Collaboration} {et~al.}(2020){Planck Collaboration}, {Aghanim}, {Akrami}, {Ashdown}, {Aumont}, {Baccigalupi}, {Ballardini}, {Banday}, {Barreiro}, {Bartolo}, {Basak}, {Battye}, {Benabed}, {Bernard}, {Bersanelli}, {Bielewicz}, {Bock}, {Bond}, {Borrill}, {Bouchet}, {Boulanger}, {Bucher}, {Burigana}, {Butler}, {Calabrese}, {Cardoso}, {Carron}, {Challinor}, {Chiang}, {Chluba}, {Colombo}, {Combet}, {Contreras}, {Crill}, {Cuttaia}, {de Bernardis}, {de Zotti}, {Delabrouille}, {Delouis}, {Di Valentino}, {Diego}, {Dor{\'e}}, {Douspis}, {Ducout}, {Dupac}, {Dusini}, {Efstathiou}, {Elsner}, {En{\ss}lin}, {Eriksen}, {Fantaye}, {Farhang}, {Fergusson}, {Fernandez-Cobos}, {Finelli}, {Forastieri}, {Frailis}, {Fraisse}, {Franceschi}, {Frolov}, {Galeotta}, {Galli}, {Ganga}, {G{\'e}nova-Santos}, {Gerbino}, {Ghosh}, {Gonz{\'a}lez-Nuevo}, {G{\'o}rski}, {Gratton}, {Gruppuso}, {Gudmundsson}, {Hamann}, {Handley}, {Hansen}, {Herranz}, {Hildebrandt}, {Hivon}, {Huang}, {Jaffe}, {Jones}, {Karakci}, {Keih{\"a}nen},
  {Keskitalo}, {Kiiveri}, {Kim}, {Kisner}, {Knox}, {Krachmalnicoff}, {Kunz}, {Kurki-Suonio}, {Lagache}, {Lamarre}, {Lasenby}, {Lattanzi}, {Lawrence}, {Le Jeune}, {Lemos}, {Lesgourgues}, {Levrier}, {Lewis}, {Liguori}, {Lilje}, {Lilley}, {Lindholm}, {L{\'o}pez-Caniego}, {Lubin}, {Ma}, {Mac{\'\i}as-P{\'e}rez}, {Maggio}, {Maino}, {Mandolesi}, {Mangilli}, {Marcos-Caballero}, {Maris}, {Martin}, {Martinelli}, {Mart{\'\i}nez-Gonz{\'a}lez}, {Matarrese}, {Mauri}, {McEwen}, {Meinhold}, {Melchiorri}, {Mennella}, {Migliaccio}, {Millea}, {Mitra}, {Miville-Desch{\^e}nes}, {Molinari}, {Montier}, {Morgante}, {Moss}, {Natoli}, {N{\o}rgaard-Nielsen}, {Pagano}, {Paoletti}, {Partridge}, {Patanchon}, {Peiris}, {Perrotta}, {Pettorino}, {Piacentini}, {Polastri}, {Polenta}, {Puget}, {Rachen}, {Reinecke}, {Remazeilles}, {Renzi}, {Rocha}, {Rosset}, {Roudier}, {Rubi{\~n}o-Mart{\'\i}n}, {Ruiz-Granados}, {Salvati}, {Sandri}, {Savelainen}, {Scott}, {Shellard}, {Sirignano}, {Sirri}, {Spencer}, {Sunyaev}, {Suur-Uski}, {Tauber}, {Tavagnacco},
  {Tenti}, {Toffolatti}, {Tomasi}, {Trombetti}, {Valenziano}, {Valiviita}, {Van Tent}, {Vibert}, {Vielva}, {Villa}, {Vittorio}, {Wandelt}, {Wehus}, {White}, {White}, {Zacchei}, \& {Zonca}}]{planck2018}
{Planck Collaboration}, {Aghanim}, N., {Akrami}, Y., {et~al.} 2020, \aap, 641, A6, \dodoi{10.1051/0004-6361/201833910}

\bibitem[{Rupke \& Veilleux(2011)}]{rupke2011integral}
Rupke, D.~S., \& Veilleux, S. 2011, The Astrophysical Journal Letters, 729, L27

\bibitem[{Rupke \& Veilleux(2013)}]{rupke2013multiphase}
---. 2013, The Astrophysical Journal, 768, 75

\bibitem[{Savage \& Sembach(1991)}]{savage1991analysis}
Savage, B.~D., \& Sembach, K.~R. 1991, The Astrophysical Journal, 379, 245

\bibitem[{Sharma {et~al.}(2025)Sharma, Arav, Korista, Bautista, Dehghanian, Byun, Walker, \& Mintz}]{Sharma2024a}
Sharma, M., Arav, N., Korista, K.~T., {et~al.} 2025, Astronomy \& Astrophysics, 693, A254

\bibitem[{Shen {et~al.}(2023)Shen, Liu, He, Zakamska, Glikman, Greene, Hu, Mou, Wylezalek, \& Rupke}]{shen2023discovery}
Shen, L., Liu, G., He, Z., {et~al.} 2023, Science Advances, 9, eadg8287

\bibitem[{Silk \& Mamon(2012)}]{silk2012current}
Silk, J., \& Mamon, G.~A. 2012, Research in Astronomy and Astrophysics, 12, 917

\bibitem[{Stockton \& Lynds(1966)}]{stockton1966remarkable}
Stockton, A., \& Lynds, C. 1966, Astrophysical Journal, vol. 144, p. 451, 144, 451

\bibitem[{Travascio {et~al.}(2024)Travascio, Piconcelli, Bischetti, Cresci, Feruglio, Perna, Vietri, Carniani, Cantalupo, Cicone, {et~al.}}]{travascio2024muse}
Travascio, A., Piconcelli, E., Bischetti, M., {et~al.} 2024, Astronomy \& Astrophysics, 686, A250

\bibitem[{Tytler \& Fan(1992)}]{tytler1992systematic}
Tytler, D., \& Fan, X.-M. 1992, Astrophysical Journal Supplement Series (ISSN 0067-0049), vol. 79, March 1992, p. 1-36., 79, 1

\bibitem[{{\"U}bler {et~al.}(2024){\"U}bler, D’Eugenio, Perna, Arribas, Jones, Bunker, Carniani, Charlot, Maiolino, del Pino, {et~al.}}]{ubler2024ga}
{\"U}bler, H., D’Eugenio, F., Perna, M., {et~al.} 2024, Monthly Notices of the Royal Astronomical Society, stae1993

\bibitem[{Vernet {et~al.}(2011)Vernet, Dekker, d’Odorico, Kaper, Kjaergaard, Hammer, Randich, Zerbi, Groot, Hjorth, {et~al.}}]{vernet2011x}
Vernet, J., Dekker, H., d’Odorico, S., {et~al.} 2011, Astronomy \& Astrophysics, 536, A105

\bibitem[{Walker {et~al.}(2022)Walker, Arav, \& Byun}]{walker2022high}
Walker, A., Arav, N., \& Byun, D. 2022, Monthly Notices of the Royal Astronomical Society, 516, 3778

\bibitem[{Williams {et~al.}(1975)Williams, Strittmatter, Carswell, \& Craine}]{williams1975splitting}
Williams, R., Strittmatter, P., Carswell, R., \& Craine, E. 1975, Astrophysical Journal, vol. 202, Dec. 1, 1975, pt. 1, p. 296-302., 202, 296

\bibitem[{Wylezalek {et~al.}(2020)Wylezalek, Flores, Zakamska, Greene, \& Riffel}]{wylezalek2020ionized}
Wylezalek, D., Flores, A.~M., Zakamska, N.~L., Greene, J.~E., \& Riffel, R.~A. 2020, Monthly Notices of the Royal Astronomical Society, 492, 4680

\bibitem[{Xu {et~al.}(2018)Xu, Arav, Miller, \& Benn}]{xu2018mini}
Xu, X., Arav, N., Miller, T., \& Benn, C. 2018, The Astrophysical Journal, 858, 39

\bibitem[{Xu {et~al.}(2019)Xu, Arav, Miller, \& Benn}]{xu2019vlt}
---. 2019, The Astrophysical Journal, 876, 105

\bibitem[{Xu {et~al.}(2021)Xu, Arav, Miller, Korista, \& Benn}]{xu2021physical}
Xu, X., Arav, N., Miller, T., Korista, K.~T., \& Benn, C. 2021, Monthly Notices of the Royal Astronomical Society, 506, 2725

\bibitem[{Xu {et~al.}(2020)Xu, Arav, Miller, Kriss, \& Plesha}]{xu2020hst}
Xu, X., Arav, N., Miller, T., Kriss, G.~A., \& Plesha, R. 2020, The Astrophysical Journal Supplement Series, 247, 42

\bibitem[{Zhao {et~al.}(\noop{3001}2025, in press.)Zhao, Sun, Liu, Wang, Arav, \& Benn}]{zhao3c191}
Zhao, Q., Sun, L., Liu, G., {et~al.} \noop{3001}2025, in press.

\bibitem[{Zhao \& Wang(2023)}]{zhao2023discovery}
Zhao, Q., \& Wang, J. 2023, The Astrophysical Journal Letters, 943, L25

\end{thebibliography}
\bibliographystyle{aasjournal}



\appendix

\section{Multiple Component Modeling of the Absorption Troughs} \label{appendix}

While determining the column densities for the outflowing system in section \ref{sec:colden}, we model the troughs with a single Gaussian. However, as Fig. \ref{fig:Al} and \ref{fig:siII} show, the absorption troughs of \ion{Al}{III}, \ion{Al}{II} and \ion{Si}{II} show additional absorption in their wings. These components are not the focus of our analysis in this work as they do not show corresponding absorption for \ion{Si}{II}*. However, to ensure that they do not affect our column density measurement for the main component, we perform a multiple component Gaussian fit for the entire absorption feature. We detail the fitting procedure for \ion{Al}{II/III} and \ion{Si}{II} separately below. 

\subsection{Details of the Models.}

\subsubsection{\ion{Al}{}}

\begin{figure}[h]
\centering
\resizebox{\hsize}{!}
{\includegraphics[width=1.20\linewidth]{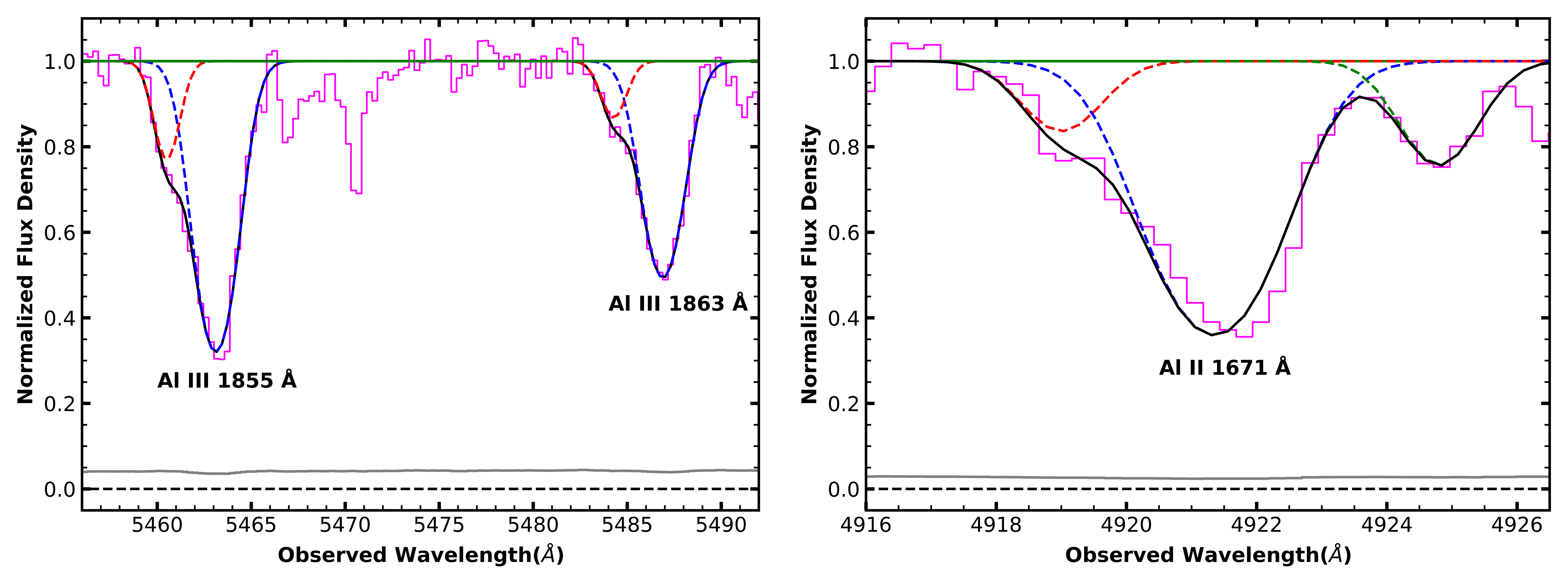}}
\caption{Multiple component Gaussian modeling for the \ion{Al}{} transitions. The main outflow component is shown in blue with the secondary component shown in red and green (for \ion{Al}{II}). The black curve shows the combined model for the troughs. The solid green lines represents the local continuum model. The gray line shows the noise in the VLT/X-shooter flux. \label{fig:Al_2G}}
\end{figure}

We first model the \ion{Al}{III} 1863 \r{A} trough with two independent Gaussian profiles for optical depth. We then use this model as a template for the \ion{Al}{III} 1855 \r{A} and \ion{Al}{II} 1671 \r{A} troughs by fixing the centroid and width of both Gaussians while allowing their depth to vary independently. In the case of \ion{Al}{II} 1671 \r{A}, we allow for an additional component to model the absorption blend in the red wing. The results for the best-fit model templates (determined by non-linear least squares) for these troughs are shown in Fig. \ref{fig:Al_2G}.

\subsubsection{\ion{Si}{II}}

\begin{figure}[h]
\centering
{\includegraphics[width=0.5\linewidth]{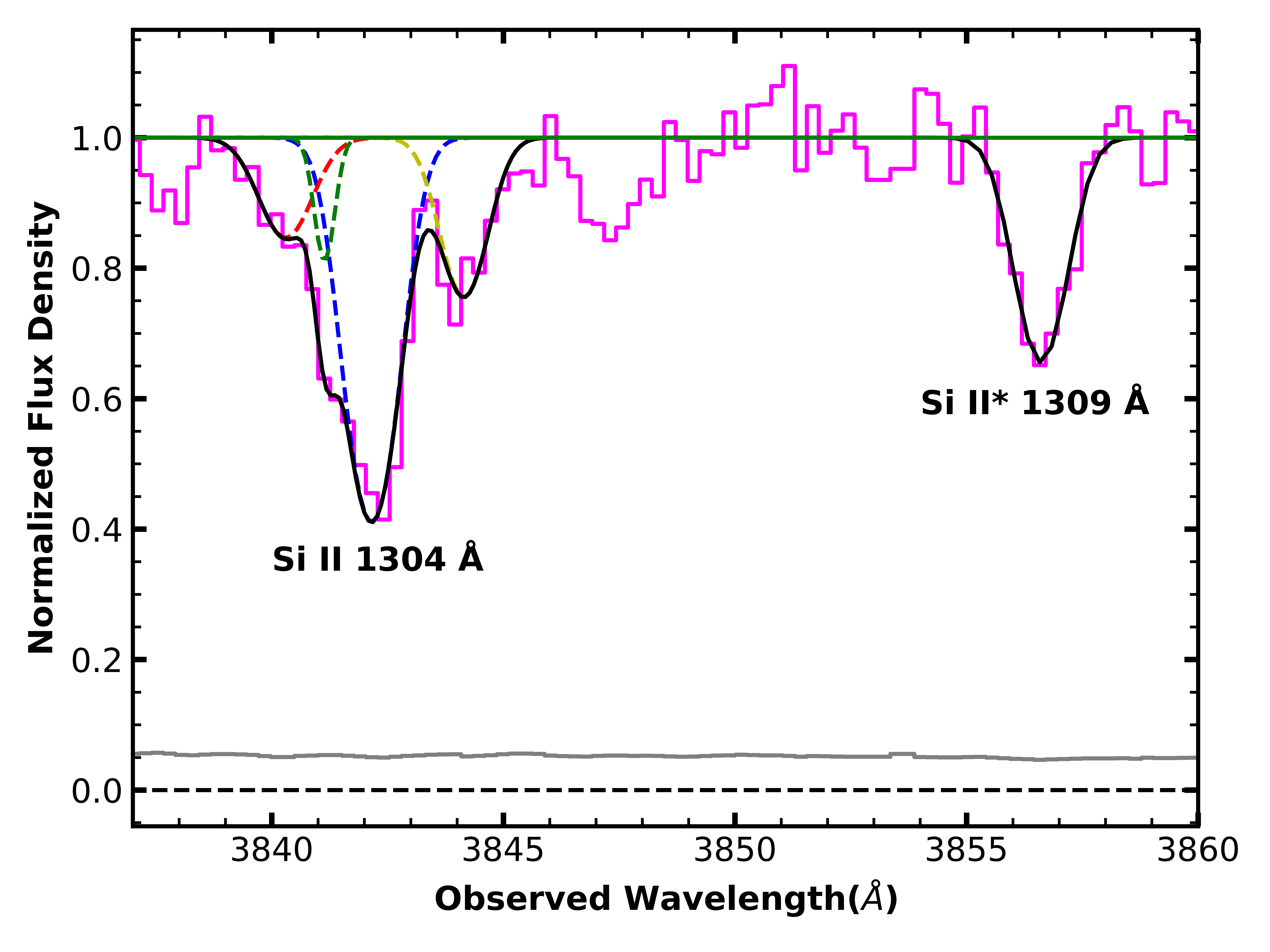}}
\caption{Multiple component Gaussian modeling for the \ion{Si}{II} transitions. The main outflow component is shown in blue with the secondary components shown in red, green and yellow. The black curve shows the combined model for the troughs. The solid green lines represents the local continuum model. The gray line shows the noise in the VLT/X-shooter flux. \label{fig:Si_3G}}
\end{figure}

The absorption trough for the \ion{Si}{II}* 1309 \r{A} transition is well modeled by a single Gaussian component as shown in Fig. \ref{fig:siII}. However the  \ion{Si}{II} 1304 \r{A} trough shows evidence for multiple components (four in particular, as can be noted by the two inflection points on the blue wing, and one more on the red wing). Therefore, we model the \ion{Si}{II} trough with four Gaussian components, of which one has fixed centroid and width based on the \ion{Si}{II}* 1309 profile. The other three components are allowed to have centroids and widths that vary independently, and the depths of all the four components are allowed to vary as well. The results for the best-fit model templates (determined by non-linear least squares) for these troughs are shown in Fig. \ref{fig:Si_3G}.

\subsection{Change in Column Density}

We now investigate the effect of the different models on the column density measurement for our main outflow component. To do so, we obtain the column density of the main component (shown in blue in Figs. \ref{fig:Al_2G} and \ref{fig:Si_3G}) from the multi component modeling and compare them with the column density obtained from the single Gaussian fitting shown in Figs. \ref{fig:Al} and \ref{fig:siII}. We report the \% change in the column density measurement due to the multiple component approach in Table \ref{table:model_comp}. The difference for all three ions are less than 5\% and therefore we use the one component Gaussian model for our analysis as it has fewer free parameters. We include this uncertainty due to our choice of models in the column densities reported in Table \ref{table:colden}.

\begingroup
\setlength{\tabcolsep}{12pt}
\renewcommand{\arraystretch}{1.5}
\vspace{-4.2cm}
\begin{table}
\caption{Difference in Column Density Measurements based on the two approaches \label{table:model_comp}}
\centering
\begin{tabular}{lc}
\hline
Ion        & $\Delta N_{ion}$ (\%)\\
\hline
\ion{Al}{ii} & -4.5 \\
\ion{Al}{iii} & -4.5 \\
\ion{Si}{ii} & -2.7 \\
\hline
\end{tabular}
\end{table}
\endgroup

\end{document}